\documentclass[times,twocolumn,final]{elsarticle}
\usepackage{graphicx}
\usepackage{multicol}
\usepackage{multirow}
\usepackage{rotating}
\usepackage{amsmath}
\DeclareMathOperator*{\argmin}{argmin}
\usepackage{amsmath}
\usepackage{xspace}
\newcommand{\name}[0]{{\sc ImplicitStainer}\xspace}

\usepackage{xcolor}
\usepackage{booktabs}
\usepackage{amssymb}
\usepackage{array}
\usepackage{adjustbox}
\usepackage{pdflscape}
\usepackage{makecell}
 \usepackage[pagewise]{lineno}
  
\newcommand{\mstd}[2]{$#1 \pm #2$}
\newcommand{\bmstd}[2]{$\mathbf{#1 \pm #2}$}

\newcommand{\cmark}{\(\checkmark\)}
\newcommand{\xmark}{\(\times\)}

\usepackage{medima}
\usepackage{framed,multirow}

\usepackage{amssymb}
\usepackage{latexsym}

\usepackage{url}
\usepackage{xcolor}
\usepackage[labelfont=bf,textfont=normalfont]{caption}
\usepackage[unicode]{hyperref}

\definecolor{newcolor}{rgb}{.8,.349,.1}

\journal{Medical Image Analysis}

\providecommand{\comma}{,}
\begin{document}

\verso{Tushar Kataria \textit{et~al.}}

\begin{frontmatter}

\title{\name: Resolution Flexible Data-Efficient Virtual Staining Using Neural Implicit Functions}

\author[1]{Tushar \snm{Kataria}\corref{cor1}}
\ead{tushar.kataria@utah.edu}
\author[2]{Beatrice \snm{Knudsen}}
\ead{beatrice.knudsen@utah.edu}
\author[1]{Shireen Y. \snm{Elhabian}}
\ead{shireen@sci.utah.edu}
\cortext[cor1]{Corresponding author: 
tushar.kataria@utah.edu}
\address[1]{Scientific Computing and Imaging Institute and Kahlert School of Computing, University of Utah, Salt Lake City-84112, USA}
\address[2]{Department of Pathology, University of Utah, Salt Lake City-84112, USA}

\begin{abstract}
Hematoxylin and eosin (H\&E)-stained slides are central to cancer diagnosis and monitoring, visualizing tissue architecture and cellular morphology. However, H\&E lacks the molecular specificity needed to distinguish cell states and functional activation. Antibody-based stains, such as immunohistochemistry (IHC), are therefore required to identify specific phenotypes (e.g., CD3$^+$ T cells or HER2-positive tumor cells), but are costly, time-consuming, and not universally available. Deep learning–based image translation methods, often termed \textit{virtual staining}, offer a complementary alternative by generating virtual immunostains directly from H\&E images.
Most existing virtual staining methods are patch-based and operate at fixed resolutions, often requiring large datasets and additional super-resolution models to generate high-resolution images. Furthermore, GAN- and diffusion-based approaches rely on generative image synthesis, which can introduce variability and synthetic artifacts that may affect the structural fidelity required for clinical applications.
We propose \name, a deterministic framework that reformulates virtual staining as a continuous pixel-level translation problem. Unlike existing patch-based approaches, \name models image translation as a continuous spatial mapping using neural implicit representations. Each target-domain pixel is predicted from a high-dimensional embedding of the corresponding source-domain H\&E pixel, its local spatial neighborhood, and explicit coordinate information. \name enables resolution-flexible training and inference, improves data efficiency, and produces deterministic, reproducible outputs.
Across multiple virtual staining tasks, including H\&E-to-IHC and H\&E-to-multiplexed immunofluorescence (mIF) translation, \name demonstrates competitive performance relative to existing methods while providing a deterministic and resolution-flexible alternative to current generative approaches.
\end{abstract}
\end{frontmatter}
\section{Introduction}

Digital pathology image analysis has the potential to play a significant role in cancer diagnosis and treatment planning \cite{romero2013digital,wen2023deep}. Pathologists routinely examine cellular-level morphology and tissue architecture to assess disease severity and guide therapeutic decisions. Tissue samples are collected and processed as either fresh-frozen (OCT-embedded, snap-frozen, cryosectioned) or FFPE (formalin-fixed, paraffin-embedded, microtome-sectioned). Sections are then stained with Hematoxylin and Eosin (H\&E) to visualize nuclei and cytoplasm \cite{fischer2008hematoxylin,feldman2014tissue}. These stained slides are then examined under a microscope to identify cancerous cells, cancer growth  patterns, and other clinically relevant features. 
In a substantial fraction (10-40\%) of cases, when information from H\&E-stained images is insufficient, pathologists request additional chemical or antibody-based tissue staining \cite{naert2013utilization,ojukwu2024immunohistochemistry,al2017immunohistochemistry}. These additional stains include chemical stains such as Periodic Acid–Schiff (PAS) or Masson’s Trichrome \cite{yang2024virtual,yang2022virtual,gadermayr2018way} or  immunohistochemical (IHC) antibody stains, such as CD3 or HER2 \cite{kataria2023automating,dubey2023structural,kataria2024staindiffuser,liu2022bci,li2023adaptive,hu2025mcs}. Multi-channel immunofluorescence (mIF) stains targeting specific molecular markers can also be performed in the research space to obtain detailed information on cell phenotypes  \cite{bian2024hemit,wu2025rosie}.
These complementary stains reveal cellular differentiation and activation states that are not visible in H\&E, providing deeper insights that support accurate diagnosis. Obtaining these specialized stains requires additional time, expertise, and financial resources \cite{raab2000cost}. As a result, many laboratories cannot routinely perform them. Deep learning–based image-to-image translation models \cite{levy2021large,wu2025rosie,dubey2023structural,kataria2024staindiffuser,li2023adaptive,liu2022bci} offer a promising and cost-effective solution for virtual tissue staining. The models leverage readily available H\&E-stained images (source domain) to generate specialized virtual stains (target domain). After passing stringent clinical validation, virtual stains may provide stain-specific information without the physical costs or resource requirements associated with real staining.

Most existing virtual staining approaches operate on image patches or tiles, typically 256×256 pixels or larger, and often benefit from large and diverse datasets to learn complex mappings between source and target domains \cite{liu2022bci,bian2024hemit,li2023adaptive,abraham2023comparison}. As image resolution and tissue complexity increase, larger and more diverse datasets can become important for robust model training and high-quality image generation \cite{liu2022bci,wang2024patch}. Moreover, patch-based training is commonly performed at fixed spatial resolutions, which can limit flexibility when generating images at scales that differ from those observed during training. Achieving higher-resolution outputs often requires multi-scale training strategies or dedicated super-resolution modules, applied either jointly or as post-processing steps \cite{liu2022bci}. In clinical pathology, however, tissue assessment is inherently multi-scale: pathologists first evaluate global tissue architecture before examining fine cellular details. Consequently, resolution-flexible virtual staining models are particularly desirable. Furthermore, training at lower or mixed resolutions can provide significant computational advantages by reducing memory consumption and overall resource requirements.

State-of-the-art virtual staining methods are predominantly based on generative adversarial networks (GANs) \cite{dubey2023structural,hu2025mcs,liu2022bci}, contrastive learning frameworks \cite{park2020contrastive,li2023adaptive}, or diffusion-based models \cite{dubey2024vims,kataria2024staindiffuser,li2023bbdm,qiu2025pasb,zhao2022egsde,oh2026virtual}. GAN-based methods often benefit from large and diverse training datasets in virtual staining. Substantial variability in tissue morphology and stain appearance across laboratories, scanners, and acquisition protocols increases the complexity of learning reliable image translation mappings \cite{chai2026impact,bouteldja2022tackling,thiringer2026scanner}. Because virtual staining datasets are often relatively small, such variability can make training more challenging and may contribute to reduced generalization, mode-collapse-related effects, and generation artifacts or hallucinated structures. Diffusion-based approaches, while effective, are inherently stochastic and can yield variable outputs for the same input image across different sampling runs \cite{mounjid2021convergence,li2023bbdm,yang2023diffusion}. While such stochasticity can enhance perceptual realism in natural image synthesis, it presents challenges in medical imaging. In virtual staining, stochastic generation may lead to hallucinated structures, over-smoothing, or subtle morphological inconsistencies that can affect downstream interpretation. Consistency and structural fidelity are therefore important considerations in virtual staining. Furthermore, structures that are sparsely represented in the training data may be challenging to model accurately for both GAN- and diffusion-based methods \cite{hussain2025visgab,um2024don}. Collectively, these challenges motivate the exploration of alternative formulations for virtual staining that prioritize deterministic predictions, structural fidelity, and data efficiency.

We propose \name, a data-efficient, resolution-flexible, and fully deterministic framework for virtual staining. \name improves data efficiency by reformulating patch-based translation as a continuous pixel-wise prediction task, enabling substantially denser supervision during training. Built on implicit neural representations, \name supports inference at arbitrary resolutions without requiring multi-scale training or auxiliary super-resolution models, while remaining compatible with multi-resolution training when desired. \name adopts a regression-based formulation that predicts each target-domain pixel (e.g., IHC or mIF) directly from its corresponding source-domain (H\&E) representation, making the framework inherently deterministic. As a result, \name produces a unique and reproducible output for a given input image, avoiding variability associated with stochastic image generation. These properties make \name well suited for medical image translation tasks, where reliability, structural consistency, and reproducibility are important considerations.

Beyond its deterministic formulation, \name explicitly models spatial context to preserve structural fidelity. Many existing patch-based approaches process image regions independently, which can make preservation of fine-grained spatial relationships challenging. In contrast, \name incorporates neighborhood information to promote contextual coherence, encouraging local predictions to remain consistent with surrounding tissue structures. To capture larger anatomical patterns—such as glands, tumor regions, blood vessels, and nerves within the stroma—the architecture integrates both convolutional and transformer-based components. The convolutional backbone extracts local morphological features, while the transformer component models long-range dependencies and global tissue context. This hybrid design enables \name to balance local precision with global consistency, an important consideration for accurate virtual staining in complex digital pathology images.

Virtual staining models are commonly evaluated using image fidelity metrics such as PSNR (Peak Signal-to-Noise Ratio) and SSIM (Structural Similarity Index), as well as feature-based metrics such as FID (Fréchet Inception Distance) and KID (Kernel Inception Distance). However, these metrics may not accurately reflect staining quality or clinical utility in medical image translation tasks \cite{kataria2025building,dubey2023structural}. While manual assessment by expert pathologists remains the gold standard, it is inherently subjective, costly, and difficult to scale to large datasets. Moreover, manual evaluation is not feasible for stains whose accuracy cannot be reliably assessed without corresponding ground-truth IHC images \cite{kataria2025building}. This limitation is especially pronounced for markers such as CD3 and CD20, which distinguish T and B lymphocytes, respectively. Although lymphocytes can often be recognized on H\&E images, their immunophenotype cannot be reliably inferred without corresponding IHC staining, making visual verification of virtual staining results challenging.

In this work, we leverage the availability of paired ground-truth stains from both public and in-house datasets to develop an automated evaluation framework based on staining accuracy. For virtual IHC staining, performance is assessed by measuring agreement between predicted and ground-truth IHC-positive regions or cell objects, following established evaluation protocols \cite{kataria2023automating,dubey2023structural,dubey2024vims,kataria2025building,hu2025mcs}. For virtual mIF staining, we introduce an accuracy-based evaluation scheme that quantifies the correspondence between stain-positive pixels in generated and reference images. By combining conventional image fidelity metrics (PSNR, SSIM, FID, and KID) with segmentation-based measures of staining accuracy, our framework provides a more comprehensive and automated assessment of virtual staining performance.

The main contributions are:
\begin{itemize}
    \item \name is a coordinate-conditioned, pixel-level image translation framework based on neural implicit representations, enabling resolution-flexible image translation and inference.

\item \name is formulated as a regression problem, making it deterministic and reproducible. Unlike GAN- and diffusion-based virtual staining approaches, \name produces a unique pixel level output for a given input image, reducing the potential for introducing synthetic structures during image generation.

\item Comprehensive ablation studies evaluate the robustness of \name across varying dataset sizes, network architectures (convolutional, transformer-based, and hybrid), image resolutions, loss-function hyperparameters, and positional encoding strategies.

\item Comprehensive qualitative and quantitative evaluations benchmark \name against more than twenty baselines across multiple virtual staining applications, including H\&E $\rightarrow$ mIF (HEMIT~\cite{bian2024hemit}) and H\&E $\rightarrow$ IHC staining tasks involving CD3 and CK8/18.

\item We employ staining-accuracy–based metrics to evaluate HEMIT~\cite{bian2024hemit} and the CD3 and CK8/18 datasets, enabling automated, objective, and interpretable assessment of virtual staining quality without requiring manual annotations.\href{https://github.com/tushaarkataria/ImplicitStainer}{Code and Additional Results.}
\end{itemize}

\section{Related Works}
\label{sec:formatting}

\textbf{Background.} Image-to-image (I2I) translation models enable the transformation of images from one domain (source domain; H\&E) to another (target domain; IHC or mIF).  Depending on the availability of corresponding image pairs, I2I translation models can be broadly divided into two main categories: \textit{(1) Paired Domain Translation (PDT)}: This approach requires pixel-wise correspondences (i.e., matched image pairs) between image patches of both source and target domains. Notable examples include Pix2Pix \cite{zhu2017unpaired}, PyramidPix2Pix \cite{liu2022bci}, BBDM \cite{li2023bbdm}, and others \cite{kataria2024staindiffuser, hu2025mcs,hussain2026vit,bian2024hemit}. \textit{(2) Unpaired Domain Translation (UDT)}: unpaired image-to-image translation methods learn mappings by matching data distributions across domains without requiring image correspondences, thereby relaxing the need for paired data acquisition. Examples of these architectures include CycleGAN \cite{zhu2017unpaired}, UNIT \cite{hu2021unit}, UGATIT \cite{kim2019u}, UVCGAN \cite{torbunov2023uvcgan}, QS-GAN\cite{hu2022qs}, AttentionGAN \cite{tang2021attentiongan} and others \cite {qiu2025pasb,park2020contrastive,chen2022eccv}.

Paired and unpaired image-to-image translation methods offer complementary strengths and weaknesses. Unpaired domain translation (UDT) greatly expands the number of possible training samples—from $\mathcal{O}(N)$ in the paired setting to 
$O(N^2)$—by allowing arbitrary pairings between images from both domains. However, UDT typically relies on distribution- or texture-level alignment rather than pixel-wise supervision, which can result in the loss of domain-specific details and hallucinated artifacts, especially in high-frequency regions \cite{wu2024stegogan,chu2017cyclegan,dubey2023structural}. Moreover, UDT models are computationally expensive due to multiple generators, discriminators, and cycle-consistency constraints. In contrast, PDT learns a direct mapping with patch-level supervision, converges faster, and is generally less prone to structural inconsistencies or artifacts, but requires paired training data that can be expensive and time-consuming to acquire. These trade-offs limit the applicability of existing approaches to medical image translation tasks, where data are scarce and structural fidelity is critical \cite{dubey2023structural,kataria2024staindiffuser,liu2022bci,li2023adaptive}. The formulation of \name assumes paired domain translation with pixel-aligned data between the two domains, but in experiments we compare against both PDT and UDT approaches using GAN- and diffusion-based models. These differences correspond to the \textit{Paired Training} dimension summarized in Table \ref{tab:design_property_comparison}.

\begin{table*}[t]
\centering
\scriptsize
\setlength{\tabcolsep}{2.2pt}
\renewcommand{\arraystretch}{1.2}
\caption{\textbf{Design Space Comparison of Virtual Staining Models.} Comparison of key design properties across representative image translation methods used for virtual staining. Existing approaches are predominantly based on patch-level generative image translation frameworks, including GAN-, contrastive-, and diffusion-based models, and vary in their use of local (convolutional) and global (transformer-based) context modeling. In contrast, \name formulates virtual staining as a deterministic, coordinate-conditioned, pixel-level neural implicit regression framework that supports resolution-flexible inference while jointly modeling local and global context.}
\label{tab:design_property_comparison}
\begin{adjustbox}{max width=\textwidth}
\begin{tabular}{p{1.4cm} p{9.4cm} c c c c c c c}
\toprule
\textbf{\makecell[l]{Method / \\ Family}} &
\textbf{Examples} &
\makecell{\textbf{Paired}\\\textbf{Training}} &
\makecell{\textbf{Neural}\\\textbf{Implicit}} &
\makecell{\textbf{Resolution}\\\textbf{Flexible}\\\textbf{Inference}} &
\makecell{\textbf{Pixel-wise}\\\textbf{Regression}} &
\makecell{\textbf{Local}\\\textbf{Context}} &
\makecell{\textbf{Global}\\\textbf{Context}} \\
\midrule

\makecell[l]{Paired \\ GAN  \\ translation} &
\makecell[l]{\textcolor{blue}{Pix2Pix}~\cite{isola2017image},
\textcolor{blue}{PyramidPix2Pix}~\cite{liu2022bci},  \textcolor{blue}{PPT} \cite{zhang2024high}, \\ 
\textcolor{blue}{M2PL-GAN} \cite{liu2026m2pl}, PSPStain~\cite{chen2024pathological}, \\ 
\textcolor{blue}{MCS-Stain}~\cite{hu2025mcs}, ASAP-Net~\cite{shaham2021spatially}} &
\cmark & \xmark & \xmark & \xmark & \cmark & \xmark \\
 &
\textcolor{red}{ViT-Stain}~\cite{hussain2026vit} &
\cmark & \xmark & \xmark & \xmark & \xmark & \cmark \\
 &
\textcolor{blue}{DualBranch}~\cite{bian2024hemit} &
\cmark & \xmark & \xmark & \xmark & \cmark & \cmark \\
\hline
 &
\textcolor{blue}{AdaptiveNCE}~\cite{li2023adaptive} &
\cmark & \xmark & \xmark & \xmark & \cmark & \xmark \\

\makecell[l]{Patch  \\Contrastive \\ translation}  &
\makecell[l]{\textcolor{blue}{CUT}~\cite{park2020contrastive},
\textcolor{blue}{FastCUT}~\cite{park2020contrastive}} &
\xmark & \xmark & \xmark & \xmark & \cmark & \xmark \\
\hline

\makecell[l]{Unpaired \\ GAN  \\ translation} &
\makecell[l]{\textcolor{blue}{CycleGAN}~\cite{zhao2020unpaired}, \textcolor{blue}{UNIT}~\cite{zhao2020unpaired},
\textcolor{blue}{UGATIT}~\cite{kim2019u},\\
\textcolor{blue}{AttentionGAN}~\cite{tang2021attentiongan},
\textcolor{blue}{QS-GAN}~\cite{hu2022qs},
\textcolor{blue}{StegoGAN}~\cite{wu2024stegogan} \\
\textcolor{blue}{SANTA}~\cite{xie2023unpaired}, Decent\cite{xieunsupervised} } &
\xmark & \xmark & \xmark & \xmark & \cmark & \xmark \\

 &
\textcolor{blue}{UVCGAN}~\cite{torbunov2023uvcgan}, \textcolor{blue}{VQ-I2I}\cite{chen2022eccv} &
\xmark & \xmark & \xmark & \xmark & \cmark & \cmark \\

\hline
\makecell[l]{Diffusion \\ / Bridge \\models} &
\makecell[l]{\textcolor{blue}{BBDM}~\cite{li2023bbdm}, \textcolor{blue}{LBBDM}~\cite{li2023bbdm}, \textcolor{red}{DiffVS}~\cite{oh2026virtual}, }
& \cmark & \xmark & \xmark & \xmark & \cmark & \xmark \\
& \makecell[l]{\textcolor{blue}{CycleDiff} ~\cite{wu2022unifying}, \textcolor{blue}{EDSDE}~\cite{zhao2022egsde}, \textcolor{blue}{UNSB}~\cite{kim2023unpaired},\\
  \textcolor{red}{PASB} \cite{qiu2025pasb}, StainFuser~\cite{jewsbury2024stainfuser}}
&
\xmark & \xmark & \xmark & \xmark & \cmark & \xmark \\
\hline
\textbf{Proposed} &
\name &
\cmark & \cmark & \cmark & \cmark & \cmark & \cmark \\

\bottomrule
\end{tabular}
\end{adjustbox}

\vspace{2pt}
\footnotesize{
\textbf{Legend:} \cmark = yes, \xmark = no.
Local Context typically corresponds to convolutional inductive bias, while Global Context corresponds to transformer-style long-range dependency modeling. \textcolor{blue}{Blue} shows the models used for comparison. \textcolor{red}{Red} Code not available or incomplete. 
}
\end{table*}

\textbf{Virtual Staining Models.} Most image-to-image (I2I) translation tasks in medical imaging—including virtual staining, MRI modality translation, and background removal—have traditionally relied on GANs or diffusion models. GAN-based approaches can be susceptible to training instability, limited output diversity, and mode-collapse-related effects, particularly when training data are limited or insufficiently diverse \cite{bau2019seeing,karras2020training}. Diffusion models generally benefit from large and diverse training datasets and often incur substantial computational cost during both training and inference \cite{dhariwal2021diffusion,xiao2021tackling,durall2020combating,thanh2020catastrophic,zhao2018bias,yang2019diversity}. Because both classes of models rely on generative image synthesis, they may introduce synthetic structures or artifacts that are not directly supported by the input image, raising concerns about reliability in safety-critical medical applications such as virtual staining \cite{kim2024tackling,wei2025responsible,oorloff2025mitigating}. Despite these limitations, most existing virtual staining methods adopt adversarial or generative image-to-image translation frameworks, such as Pix2Pix, CycleGAN and more recently diffusion-based models across a wide range of virtual staining tasks: H\&E $\rightarrow$ immunofluorescence \cite{bian2024hemit,wu2025rosie}, immunofluorescence $\rightarrow$ H\&E \cite{rivenson2019phasestain}, H\&E $\rightarrow$ IHC \cite{li2023adaptive,liu2022bci,dubey2023structural,kataria2024staindiffuser,hu2025mcs,qiu2025pasb}, FFPE $\rightarrow$ H\&E \cite{ozyoruk2022deep,ho2024f2fldm,ho2024disc}, and chemical stain synthesis \cite{yang2024virtual,gadermayr2018way,yang2022virtual}. These approaches are often enhanced with pathology-specific priors—such as multi-resolution fusion \cite{liu2022bci}, contrastive patch losses \cite{li2023adaptive}, structural constraints \cite{dubey2023structural}, stain-channel supervision \cite{klockner2025h}, and geometry-aware regularization \cite{peng2024advancing}. However, such augmentations do not fundamentally address the inherent stochasticity and artifact generation risks of GAN- and diffusion-based models. This limitation directly relates to the \textit{pixel wise regression} property in Table \ref{tab:design_property_comparison}, as most existing approaches rely on generative image synthesis rather than direct pixel regression.

We propose \name, which reframes image domain translation as a deterministic, pixel-level regression problem, enabling precise, context-aware, and structurally faithful predictions. While regression-based I2I methods have been explored in related tasks such as depth estimation and segmentation \cite{dieckhaus2022logistic}, these approaches typically operate at the patch level. To the best of our knowledge, direct pixel-wise regression has not been explored for virtual staining. The closest architecturally related method is ASAP-Net \cite{shaham2021spatially}, which employs per-pixel MLPs combined with convolutional features for fast, high-resolution GAN-based synthesis. However, ASAP-Net is primarily designed for GAN-based image synthesis, and its performance has been reported to be comparable to that of Pix2Pix-style models \cite{isola2017image}. Unlike ASAP-Net, \name uses a single shared MLP with a hybrid convolutional–transformer backbone, explicitly designed for data-efficient, deterministic virtual staining while reducing the potential for hallucinated structures. Table \ref{tab:design_property_comparison} summarizes how our proposed \name occupies a distinct point in the design space and unifies key design properties—such as determinism, resolution flexibility, and joint local–global context modeling—that are not simultaneously achieved by prior methods.

{\textbf{Convolution-Transformer Backbones.} Histopathology and medical imaging patterns span multiple spatial scales. Convolutional neural networks effectively capture local morphological features due to their inductive bias, while transformer-based architectures excel at modeling long-range dependencies and global context \cite{raghu2021vision,ding2022scaling,hatamizadeh2023global}. This complementary behavior has motivated hybrid architectures that combine convolutional and transformer components across diverse vision tasks \cite{liu2022convnet,oquab2023dinov2,carion2020end,liu2023hybrid}. In medical imaging, similar approaches have been explored in semi-supervised segmentation via cross-teaching frameworks \cite{luo2022semi}, conditional image generation \cite{chen2022eccv,torbunov2023uvcgan}, and virtual staining through DualBranch Pix2Pix \cite{bian2024hemit}, which integrates CNN and transformer features within a GAN-based generator. However, these methods are generally designed within generative image translation frameworks and are not formulated as deterministic pixel-level regression models. As summarized in Table \ref{tab:design_property_comparison}, relatively few virtual staining methods jointly leverage local morphological features and global contextual information. In contrast, \name jointly models both within a deterministic, pixel-level framework. To assess the impact of these design choices, we include representative methods from these categories in our comparative evaluation.

\textbf{Applications of Neural Implicit Functions.} Neural implicit functions were originally introduced to represent 3D geometry, including meshes, point clouds, and surfaces, by learning continuous occupancy or signed distance fields conditioned on spatial coordinates \cite{sitzmann2019scene,vetsch2022neuralmeshing,fujiwara2020neural,sitzmann2020implicit,mescheder2019occupancy,wang2021neural,huang2024surface}. Early methods modeled individual shapes with separate implicit functions, while later approaches introduced latent codes to enable representation of multiple objects within a shared network \cite{park2019deepsdf}.

Neural implicit representations are powerful because they operate on a continuous spatial domain, enabling flexible modeling across varying resolutions and spatial locations. More recently, they have been explored for continuous super-resolution \cite{chen2021learning,shen2025fast} and image generation \cite{anokhin2021image}, demonstrating promising performance across a range of vision tasks. Motivated by their continuous representation capability, we formulate domain translation as a pixel-level translation problem using neural implicit functions. Our approach shares conceptual similarities with LIIF \cite{chen2021learning}, but differs in several key respects. LIIF is designed for super-resolution, where the objective is to reconstruct high-resolution signals from low-resolution inputs within the same domain. In contrast, our method addresses cross-domain pixel-level translation, requiring the learning of semantic correspondences between distinct domains while preserving structural consistency. Architecturally, LIIF relies on a purely convolutional backbone and conditions predictions on local features, whereas our framework integrates convolutional and transformer encoders to jointly capture local morphology and global tissue context. In addition, unlike LIIF's reliance on relative coordinate encoding, we incorporate learnable positional embeddings to encode richer spatial priors. Finally, while LIIF primarily targets resolution-agnostic inference, our framework supports both multi-resolution training and inference, enabling scalable performance across varying image resolutions while maintaining a deterministic prediction objective.

Collectively, these design choices extend neural implicit representations beyond interpolation-based reconstruction to cross-domain medical image translation in \name.

\section{Method: \name Architecture}
\begin{figure*}[!thb]
    \centering
    \includegraphics[trim={1.8cm 4.55cm 2.6cm 5.0cm}, clip=true, width=1.0\linewidth]{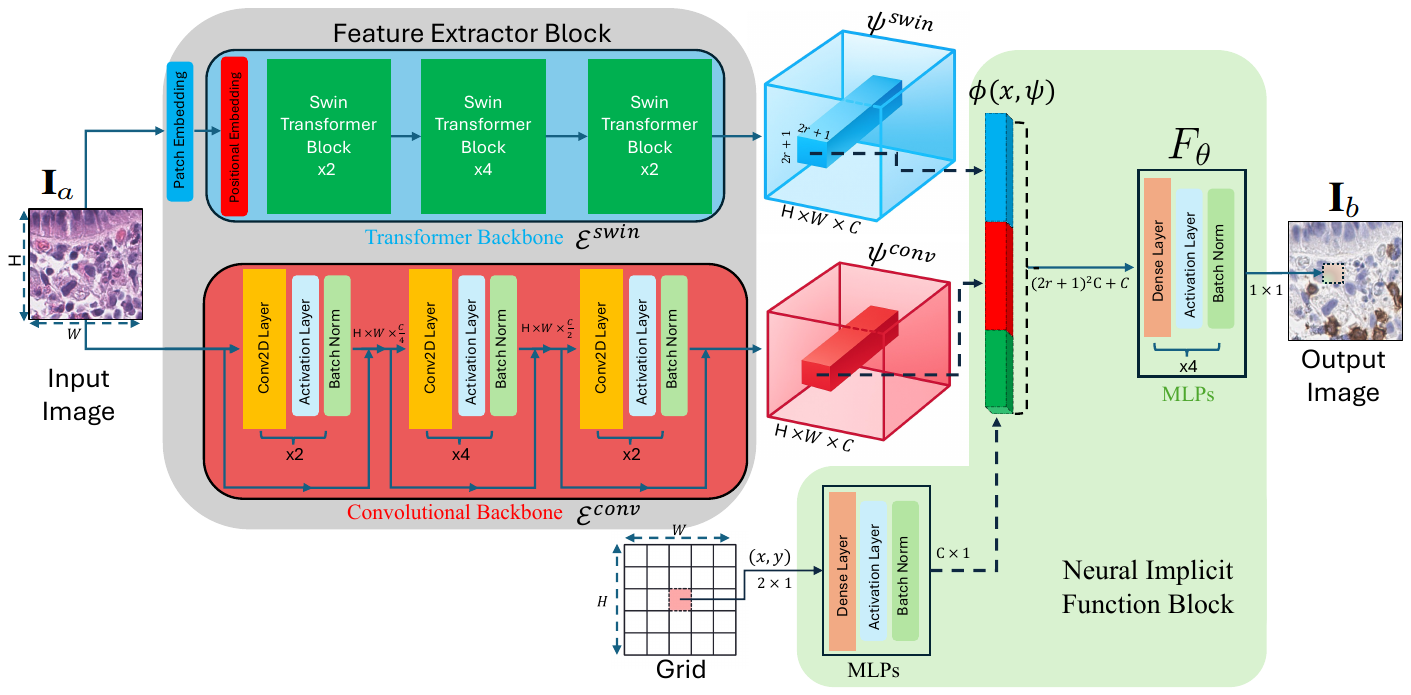}
    \caption{\textbf{\name Architecture.} \textbf{Feature Extractor Block}: The proposed model integrates convolutional and transformer backbones to learn pixel-wise representations that balance local and global contextual information. No downsampling layers are used in the feature extraction block to ensure that spatial resolution is preserved and the learned representations remain directly aligned with the input pixels for accurate pixel-level translation. \textbf{Neural Implicit Function Block}: The neural implicit function block consists of multilayer perceptrons (MLPs) with ReLU activations that map the concatenated features from the convolutional and transformer backbones, along with the grid coordinates, to output pixel values. When the entire grid is processed as a batch, the model can reconstruct the complete translated image.}
    \label{fig:pipeline_fig}
     \vspace{-1.5em}
\end{figure*}

\noindent\textbf{Patch Translation Formulation.}  
Image-to-image translation for virtual staining can be formulated as learning a function 
$G: \mathcal{A} \rightarrow \mathcal{B}$ that maps an image patch from domain $\mathcal{A}$ 
(e.g., H\&E) to domain $\mathcal{B}$ (e.g., IHC or mIF). 
During training, image patches are sampled from both domains at a fixed spatial resolution 
using a grid of size $H \times W$, where $H$ and $W$ denote the height and width of the patch.
For the paired setting, the training dataset is defined as
\[
\mathcal{D}_{train} = \{(\mathbf{I}_a^n, \mathbf{I}_b^n)\}_{n=1}^{N},
\]
where $\mathbf{I}_a$ and $\mathbf{I}_b$ denote corresponding patches sampled from 
domains $\mathcal{A}$ and $\mathcal{B}$, respectively. 
A deep neural network parameterized by $\theta$ learns the mapping $G$ by minimizing

\begin{align}\label{eq:1}
\argmin_{\theta} \mathbb{E}_{(\mathbf{I}_a,\mathbf{I}_b)\sim \mathcal{D}_{train}}
\left[ ||\mathbf{I}_b - G_{\theta}(\mathbf{I}_a)||_p \right], 
\end{align}

where $||\cdot||_p$ denotes the $L_p$ norm used for optimization. 

\vspace{0.05in}
\noindent\textbf{Proposed Continuous Translation Formulation.}
Let images from domains $\mathcal{A}$ and $\mathcal{B}$ be defined over a continuous spatial domain 
$\Omega \subset \mathbb{R}^2$, where spatial locations are denoted by coordinates 
$\mathbf{x} \in \Omega$. A source-domain image $I_a \in \mathcal{A}$ and its corresponding 
target-domain image $I_b \in \mathcal{B}$ are functions defined over $\Omega$, such that 
$I_a(\mathbf{x}) \in \mathbb{R}^c$ and $I_b(\mathbf{x}) \in \mathbb{R}^c$, where $c$ denotes the 
number of image channels (e.g., stain channels).

We define a continuous coordinate-conditioned translation function 
$F_\theta : \Omega \times \mathbb{R}^d \rightarrow \mathbb{R}^c$, parameterized by $\theta$, 
that predicts the target-domain pixel value at location $\mathbf{x}$. The function takes as input 
the spatial coordinate $\mathbf{x}$ together with a feature representation 
$\phi(\mathbf{x}, I_a) \in \mathbb{R}^d$, where $\phi$ denotes a feature extraction function that 
produces a $d$-dimensional feature vector capturing image-based contextual information from the 
source image $I_a$ conditioned on the spatial location $\mathbf{x}$.
The continuous translation function is learned by minimizing the expected pixel-wise discrepancy 
between the predicted and ground-truth target images:

\begin{align}\label{eq:2}
\argmin_{\theta} \mathbb{E}_{(I_a,I_b)\sim \mathcal{D}_{train}}
\left[
\frac{1}{|\Omega|}\int_{\Omega}
\| I_b(\mathbf{x}) - F_\theta(\mathbf{x}, \phi(\mathbf{x}, I_a)) \|_p
\, d\mathbf{x}
\right].
\end{align}

Unlike Eq.~\ref{eq:1}, which operates on discrete image patches sampled on a fixed grid, 
our formulation defines the translation function over the continuous spatial domain $\Omega$. 
In practice, the integral is approximated through discrete sampling of spatial coordinates 
during training. This continuous formulation decouples the model from a fixed patch resolution, 
enabling resolution-agnostic training and inference.

\vspace{0.05in}
\noindent\textbf{\name Architecture.}
This section describes how \name implements the proposed continuous pixel translation formulation defined in Eq.~\ref{eq:2}. 
Although the formulation is defined over the continuous spatial domain $\Omega$, in practice the domain is discretized into a square grid 
$\Omega_g = \{1,\ldots,H\} \times \{1,\ldots,W\}$, where each pixel coordinate $\mathbf{x}$ lies on this grid. 
The pixel coordinates are normalized to the range $[-1,1]$, which enables higher-resolution outputs to be generated simply by sampling spatial coordinates at finer resolutions during inference. The full architecture is shown in Figure \ref{fig:pipeline_fig}.

To compute the feature representation $\phi(\mathbf{x}, I_a)$ described in Eq.~\ref{eq:2}, we employ a dual-encoder architecture that extracts multi-scale image features from the source image $I_a \in \mathcal{A}$. 
The feature extractor consists of a convolutional encoder that captures local (short-range) dependencies and a Swin Transformer encoder that models long-range contextual dependencies. 
Unlike conventional architectures that rely on spatial downsampling via pooling in CNNs and patch merging in transformers for feature extraction, we preserve the full spatial resolution in both encoders. 
Maintaining dense feature maps ensures that fine-grained local structures and pixel-wise correspondences are preserved, which is critical for accurate image translation.
Formally, the two encoders produce feature maps
\begin{equation}
\mathbf{\Psi}^{conv},\mathbf{\Psi}^{swin} =
\mathcal{E}^{conv}(\mathbf{I}_a),\mathcal{E}^{swin}(\mathbf{I}_a)
\end{equation}
where $\mathcal{E}^{conv}$ and $\mathcal{E}^{swin}$ denote the convolutional and Swin Transformer encoders, respectively (Fig.~\ref{fig:pipeline_fig}). 
Both feature maps lie in $\mathbb{R}^{C\times H\times W}$, where $C$ is the number of output channels.
The two representations are concatenated along the channel dimension to obtain
\begin{equation}
\mathbf{\Psi} = \operatorname{concat}(\mathbf{\Psi}^{conv},\mathbf{\Psi}^{swin})
\end{equation}
resulting in a feature tensor $\mathbf{\Psi}\in\mathbb{R}^{2C\times H\times W}$. 
For notational convenience we define the joint encoder as
\begin{equation}
\mathcal{E} = (\mathcal{E}^{conv},\mathcal{E}^{swin}),
\qquad
\mathbf{\Psi} = \mathcal{E}(\mathbf{I}_a)
\end{equation}
where $\mathbf{\Psi}(\mathbf{x})$ denotes the feature representation corresponding to pixel coordinate $\mathbf{x}$.

To capture richer contextual information around each location, we construct a local neighborhood representation centered at spatial coordinate $\mathbf{x}=(x,y)$. Let $r$ denote the neighborhood radius. The local feature window is defined as
\begin{equation}
\mathbf{\Phi}(\mathbf{\Psi},\mathbf{x};r)=
\mathbf{\Psi}[x-r:x+r,\,y-r:y+r]
\end{equation}
which corresponds to a $(2r+1)\times(2r+1)$ spatial neighborhood centered at $\mathbf{x}$.
The extracted window is flattened and concatenated to form a high-dimensional vector representation. 
This vector corresponds to the feature representation $\phi(\mathbf{x},I_a)\in\mathbb{R}^d$ introduced in Eq.~\ref{eq:2}, where $d=2C(2r+1)^2$.

For resolution-agnostic inference, the feature representation $\phi(\mathbf{x},I_a)$ is obtained by sampling the neighborhood around the nearest discrete pixel $\hat{\mathbf{x}}$ corresponding to the normalized coordinate $\mathbf{x}$. 
Additionally, the coordinate $\mathbf{x}$ is projected into learnable positional embeddings that are appended to the feature vector (Fig.~\ref{fig:pipeline_fig}).

To implement the continuous translation function, we employ an implicit neural model $F_\theta$ parameterized by $\theta$. 
The function $F_\theta$ is implemented as a multi-layer perceptron (MLP) composed of fully connected layers with ReLU activations. 
Given a spatial coordinate $\mathbf{x}$ and its corresponding feature representation $\phi(\mathbf{x},I_a)$, the network predicts the pixel value in the target domain:
\begin{equation}
\widehat{\mathbf{I}}_b(\mathbf{x})
=
F_\theta(\mathbf{x},\phi(\mathbf{x},I_a))
=
F_\theta\left(
\mathbf{x},
\mathbf{\Phi}(\mathcal{E}(\mathbf{I}_a),\mathbf{x};r)
\right)
\end{equation}
where $\mathbf{x}\in\Omega_g$ and $\widehat{\mathbf{I}}_b(\mathbf{x})\in\mathbb{R}^c$ represents the predicted pixel value in the target domain.

The parameters $\theta$, $\mathcal{E}^{conv}$, and $\mathcal{E}^{swin}$ are trained end-to-end using a regression loss that minimizes the difference between the predicted and ground-truth target images:
\begin{align}
\mathcal{L}_{Imp} =
\mathbb{E}_{(\mathbf{I}_a,\mathbf{I}_b)\sim\mathcal{D}_{train}}
\left[
\frac{1}{|\Omega_g|}
\sum_{\mathbf{x}\in\Omega_g}
\|
\mathbf{I}_b(\mathbf{x})-\widehat{\mathbf{I}}_b(\mathbf{x})
\|_1
\right]
\end{align}

Minimizing only this pixel-wise regression loss often produces overly smoothed images that fail to capture high-frequency structures and texture patterns in domain $\mathcal{B}$. This limitation of pixel-wise losses such as L1 or L2 is well documented in image synthesis tasks. 
To address this issue, we incorporate a perceptual loss as a regularization term. The perceptual loss compares feature activations of the generated image $\widehat{\mathbf{I}}_b$ and the ground-truth image $\mathbf{I}_b$ in a pretrained network, encouraging the reconstructed images to not only match the pixel-wise distribution but also preserve structural and texture-based details \cite{johnson2016perceptual}. Such perceptual supervision has been widely used in super-resolution and generative modeling to enhance the visual fidelity of synthesized images \cite{zhang2018unreasonable,wang2018esrgan}. The final training objective becomes
\begin{equation}
\mathcal{L}_{total}=
\mathcal{L}_{Imp}+\lambda\mathcal{L}_{perceptual}
\end{equation}
where $\lambda$ controls the contribution of the perceptual loss.

\section{Experimental Details and Evaluation}

To comprehensively evaluate the proposed framework, we conduct experiments on two virtual staining tasks. (1) \textbf{Immunohistochemical (IHC) virtual staining}, where we utilize two internal paired datasets: virtual cytokeratin staining (H\&E $\rightarrow$ CK8/18) and virtual lymphocyte staining (H\&E $\rightarrow$ CD3). (2) \textbf{Multiplex immunofluorescence (mIF) virtual staining}, where we employ the publicly available HEMIT dataset, in which the objective is to translate an H\&E image into corresponding multi-channel immunofluorescence outputs (PanCK, CD3, and DAPI). Dataset statistics, including the number of image patches in each dataset, are summarized in Table~\ref{tab:dataset}.

All datasets used in this study consist of paired H\&E and corresponding immunostained images (IHC or mIF) acquired from the same tissue sections. Consequently, the image pairs are spatially registered and aligned, enabling supervised learning and quantitative evaluation. This alignment facilitates automated assessment of virtual staining performance using both conventional image-fidelity metrics (e.g., SSIM, PSNR, and feature-based similarity measures) and staining-accuracy–based metrics, reducing reliance on subjective manual assessment. Furthermore, the inclusion of diverse biomarkers and staining modalities enables evaluation of the proposed framework across multiple virtual staining settings.

\textbf{Private Dataset Acquisition.}
The internal dataset comprises H\&E whole-slide images (WSIs) obtained from surveillance colonoscopies of patients with active ulcerative colitis (92 tissue sections). All slides were initially stained with H\&E and scanned at 40x magnification (0.23 $\mu$m/pixel) using an Aperio AT2 scanner. The same tissue sections were subsequently restained with CD3 or CK8/18 antibodies on a Leica Bond-III automated platform. During this process, heat-induced epitope retrieval removes the original H\&E stain, enabling accurate immunohistochemical (IHC) labeling on the same tissue. Following IHC staining, slides were rescanned under identical imaging conditions and paired with their corresponding H\&E images. \footnote{The pathology archive at the University of Utah collected samples following all informed consent guidelines. All protocols were approved by the Institutional Review Board (IRB) at the University of Utah (IRB 00140202 and IRB 00057287). The study was conducted under a waiver of consent because all Health Insurance Portability and Accountability Act (HIPAA)–sensitive data were removed before slide use. No demographic or clinical information was used, and links to medical records were destroyed prior to any image processing.}

To account for spatial misalignment between scans, we performed multi-resolution image registration using ANTs \cite{kataria2023automating}. This coarse-to-fine strategy aligns tissue structures at both global and local scales. Following automated registration, all slide pairs were manually reviewed to identify residual misalignments and registration artifacts. Where necessary, manual refinements were applied to improve local alignment and achieve high-quality spatial correspondence between H\&E and IHC images. This preprocessing step is important for reliable paired training and quantitative evaluation of virtual staining methods.

\textit{In House Data Preparation for virtual staining}. {The train–test splits were constructed with strict patient- and slide-level separation, ensuring that no samples from the same patient or slide appear in both sets and thereby eliminating any risk of data leakage. This setup enforces a realistic evaluation of model generalization to unseen patients. For experiments involving reduced dataset sizes, we subsample only the training set by selecting patches from a smaller subset of patients, while keeping the test set unchanged. This preserves the diversity and consistency of the evaluation data across all experiments. As a result, all models are assessed on the exact same test set, ensuring fair and comparable evaluation under varying training data regimes.}

\textbf{Public Dataset Used.} Although controlled paired datasets for IHC virtual staining are difficult to obtain—requiring precise restaining and accurate registration of the same tissue sections—paired datasets are more readily available for multiplexed immunofluorescence (mIF). This supports our evaluation in two key ways: (1) it enables benchmarking on a public dataset and facilitates reproducibility through model release, and (2) it provides an additional domain to assess the generalization of our approach. For details on the public HEMIT dataset \cite{bian2024hemit}, we refer readers to the original publication. We follow the dataset provider’s protocol and use the official train–test split. For fair comparison across baselines, all models are trained on 256×256 patches, which are extracted from the original images using a non-overlapping sliding window without downsampling.

{The test set, for both public and private datasets, is strictly held out and never used for training or validation. We exclude test data from all stages of model development, ensuring that reported results reflect true generalization performance.}

\begin{table}[!htb]
\setlength{\tabcolsep}{8pt}
    \centering
    \scalebox{1.0}{
    \begin{tabular}{c|c|cc}
      &        & \multicolumn{2}{c}{No. Of Patches} \\
      & \bf Task    &  \it Train & \it Test \\
       \hline
       Public  &  HEMIT ( H\&E$\rightarrow$ mIF)& 55372 & 7899 \\
      \hline
      In-house & H\&E$\rightarrow$CD3    & 59362 & 6920\\
      In-house & H\&E$\rightarrow$ CK8/18 & 57887 & 6460 \\
    \end{tabular}}
    \caption{\textbf{Dataset Details for Each Virtual Stain.} We use two in-house datasets for IHC virtual staining. In addition, we utilize the publicly available HEMIT dataset \cite{bian2024hemit} for multi-channel immunofluorescence (mIF) virtual staining, which includes three channels: PanCK, CD3, and DAPI.}
    \label{tab:dataset}
\end{table}

\textbf{Baselines and Implementation Details.}We compare \name against a comprehensive suite of image-to-image (I2I) translation baselines (Table~\ref{tab:design_property_comparison}), encompassing GAN-, contrastive-, and diffusion-based architectures. Because neural implicit representations have not previously been explored for virtual staining, we benchmark against representative methods from the virtual staining and medical image translation literature, including convolution--transformer hybrids such as UVCGAN and Dual-Branch Pix2Pix. In total, we evaluate 15 unpaired and 8 paired approaches (Tables ~\ref{tab:results_main_table_CK818}, \ref{tab:results_main_table_CD3} and  \ref{tab:quantiative_metrics}). Unless otherwise specified, all baselines were trained using the authors' recommended settings under a unified protocol consisting of a batch size of 4, 200 training epochs, and 256×256 image patches.
\begin{itemize}
    \item \textbf{Paired Baselines}. We evaluate Pix2Pix~\cite{isola2017image}, PyramidPix2Pix~\cite{liu2022bci}, DualBranch Pix2Pix~\cite{bian2024hemit}, and AdaptiveNCE~\cite{li2023adaptive}. We also include diffusion-based models—EDSDE~\cite{zhao2020unpaired}, BBDM/LBBDM~\cite{li2023bbdm}, and CycleDiffusion~\cite{wu2022unifying}—though these are evaluated strictly on the IHC task due to their high computational cost. Notably, BBDM and LBBDM did not achieve stable convergence under our experimental setup on the mIF virtual staining task and are excluded from those specific comparisons. ASAP-Net~\cite{shaham2021spatially} is not included because it requires additional label supervision and was originally designed for high-resolution image synthesis rather than virtual staining. Furthermore, prior studies have reported performance comparable to Pix2Pix-style architectures on standard image translation benchmarks. To compare with the recent methods we additionally added comparisons with PPT \cite{zhang2024high}, MCS-Stain \cite{hu2025mcs} and M2PL-GAN \cite{liu2026m2pl}. 
    \item \textbf{Unpaired Baselines.} Our selection includes classic models like CycleGAN~\cite{zhu2017unpaired}, CUT/FastCUT~\cite{park2020contrastive}, and UNIT~\cite{zhu2017unpaired}; attention- and transformer-based variants such as AttentionGAN~\cite{tang2021attentiongan}, QS-GAN~\cite{hu2022qs}, UGATIT~\cite{kim2019u}, and UVCGAN~\cite{torbunov2023uvcgan}; as well as recent state-of-the-art models like UNSB~\cite{kim2023unpaired} and StegoGAN~\cite{wu2024stegogan}. Among the unpaired baselines, UVCGAN, AttentionGAN, and QS-GAN are particularly relevant comparisons, as they incorporate both convolutional and transformer-based components.
    \item \textbf{Excluded Methods.} Recent methods such as PSPStain~\cite{chen2024pathological}, Cytoland~\cite{liu2025robust}, PASB~\cite{qiu2025pasb}(UNSB \cite{kim2023unpaired} a closely related method is included instead), and VitStain~\cite{hussain2026vit} are not included in our comparison because publicly available implementations or sufficient training details were unavailable, preventing reproducible evaluation. StainFuser~\cite{jewsbury2024stainfuser} is designed for stain normalization rather than virtual stain generation; therefore, these methods fall outside the scope of our study. DiffVS \cite{oh2026virtual} when trained locally on our copy of HEMIT dataset didn't converge to a viable solution so the numbers are reported from their paper. 
\end{itemize}

We believe that the selected comparison set provides a comprehensive evaluation against both established image-to-image translation metrics and more recent adaptations of these methods for virtual staining.

\textbf{\name Training Details.} For \name, we use frozen pretrained VGG and ResNet networks to compute the perceptual loss ($\lambda = 0.1$). Models are optimized using the AdamW optimizer with a learning rate of $1 \times 10^{-4}$. The feature extraction module consists of a 12-layer convolutional encoder (3$\times$3 kernels, stride 1, channel dimensions ranging from 32 to 256) and a parallel Swin Transformer branch with an embedding dimension of 256, 8 attention heads, and a depth of 6. Features from both branches are concatenated, as illustrated in Figure~\ref{fig:pipeline_fig}. Unless otherwise specified, all \name variants use a neighborhood radius of $r = 3$. All experiments were conducted on NVIDIA H200 GPUs.

\subsection{Evaluation Protocol}

Most prior evaluations of virtual staining rely heavily on image-fidelity metrics such as FID (Fréchet Inception Distance), PSNR (Peak Signal-to-Noise Ratio), and SSIM (Structural Similarity Index). This reliance stems in part from the prevalence of unpaired datasets in the literature, where direct measurement of staining accuracy is not possible, leading researchers to use image-fidelity metrics as indirect measures of virtual staining performance. While expert review by pathologists is sometimes employed as an additional evaluation strategy, it has important limitations: pathologists cannot reliably infer specific immunohistochemical (IHC) marker expression (e.g., CD3 or CD20) from H\&E images alone without corresponding ground-truth staining~\cite{kataria2025building}. Furthermore, such evaluations are typically conducted on relatively small cohorts and often emphasize visual realism and perceptual quality rather than direct assessment of staining accuracy~\cite{liu2022bci,li2023adaptive,hu2025mcs}. Consequently, texture- and distribution-based metrics alone may not accurately reflect the biological fidelity of generated stains~\cite{kataria2025building}. 
 \begin{figure*}[!htb]
    \centering
    \includegraphics[trim={0.0cm 9cm 0.0cm 0.0cm}, clip=true, width=1.0\linewidth]{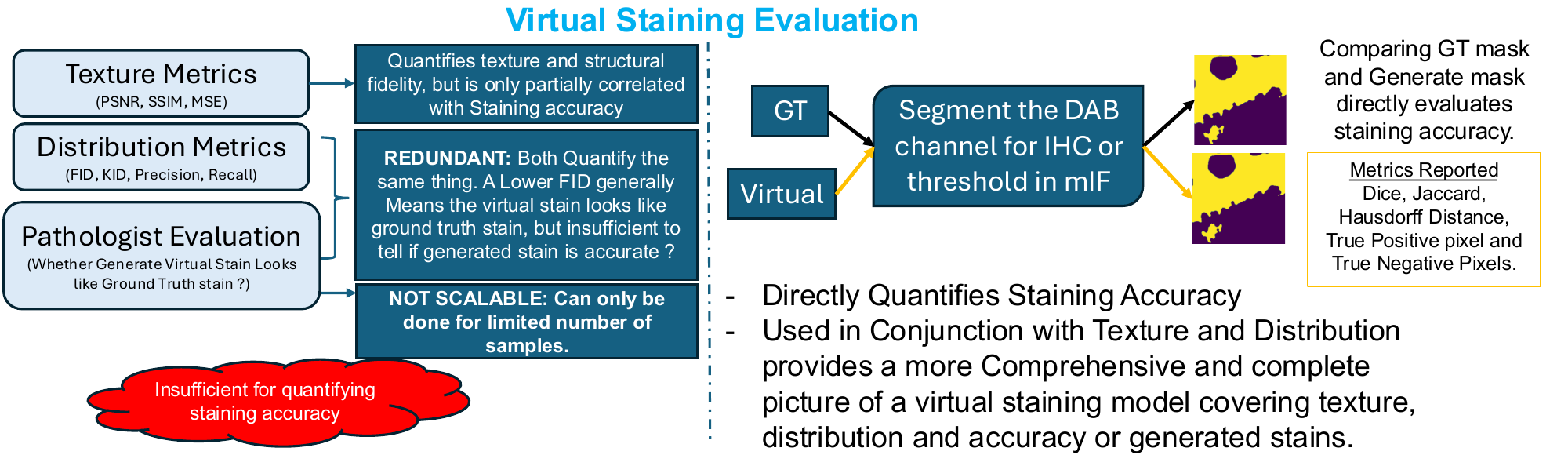}
    \caption{\textbf{Evaluation Pipeline}. Pictorial representation of our complete evaluation pipeline.}
    \label{fig:EvaluationFigure}
\end{figure*} 
To address these limitations, all datasets used in this study are paired, consisting of H\&E images and corresponding IHC or mIF images acquired from the same tissue sections, including HEMIT~\cite{bian2024hemit}. This paired structure enables direct, automated, and scalable evaluation of staining accuracy while substantially reducing reliance on subjective human assessment. Accordingly, we establish a comprehensive evaluation protocol that combines conventional image-fidelity metrics (to assess visual similarity) with segmentation-based metrics designed to quantify staining accuracy.  A pictorial representation of the complete evaluation pipeline is shown in Figure~\ref{fig:EvaluationFigure}.

\textbf{Image Realism Metrics.} For assessing the visual and cosmetic realism of the generated images, we report standard pixel-level metrics, namely Peak Signal-to-Noise Ratio (PSNR), Structural Similarity Index (SSIM), and Mean Squared Error (MSE). Both PSNR and SSIM are programmatically computed using the \texttt{skimage} library. To complement these pixel-level measures and further assess broader distributional similarity, we incorporate latent-space metrics commonly adopted within generative modeling frameworks: (a) Fréchet Inception Distance (FID)\footnote{https://github.com/mseitzer/pytorch-fid}, (b) Kernel Inception Distance (KID), alongside (c) precision and (d) recall measures\footnote{https://github.com/photosynthesis-team/piq}. These latent metrics are calculated using deep feature representations extracted from an ImageNet-pretrained network with a 2048-dimensional embedding space. Notably, we deliberately opt against utilizing pathology-specific pretrained encoders. Prior work has demonstrated that these evaluation metrics remain highly correlated regardless of whether domain-specific biomedical features or natural-image features are utilized~\cite{kataria2025building}; employing a standard ImageNet backbone thus ensures architectural consistency and direct comparability with the existing body of virtual staining literature.

\textbf{Staining Accuracy Metrics.} To accurately quantify the biological fidelity of virtual immunohistochemical (IHC) markers, we implement an established segmentation-based validation protocol~\cite{kataria2025building}. For the CK8/18 and CD3 datasets, each IHC image is first digitally deconstructed into its constituent Hematoxylin, Eosin, and DAB (brown) color components via color deconvolution. Because the isolated DAB channel specifically corresponds to the IHC-positive signal, Otsu-based segmentation was initially used to determine a suitable thresholding range. A fixed threshold of 0.035 was subsequently selected through systematic analysis of real IHC images and refined through consultation with an expert pathologist and manual verification. This threshold was then applied uniformly to both ground-truth and generated images to extract binary masks of positive pixels~\cite{kataria2023automating}.

To illustrate the threshold selection process, qualitative examples demonstrating model sensitivity across different threshold values are provided in  Supplementary Fig.~\ref{fig:Threshold_senstivity}. Lower thresholds tend to produce over-segmentation, whereas excessively high thresholds can result in under-segmentation for both the CD3 and CK8/18 markers. We observed a stable segmentation region within the 0.035–0.05 range and therefore selected 0.035 as a consistent operating point to ensure that lightly stained cellular regions were retained during evaluation. Applying the same threshold to both reference and virtually stained images ensures a consistent evaluation protocol across all methods.

The resulting binary masks—illustrated in Supplementary Figure~\ref{fig:qualitative_results_threshold}—serve as the basis for assessing staining accuracy and quantifying how well virtual staining models recover the spatial distribution of IHC-positive regions. To provide a comprehensive pixel-level evaluation of localization performance, we report standard segmentation metrics computed directly from these DAB-derived masks, including the Dice coefficient, Intersection over Union (IoU), Hausdorff Distance (HD), True Positive Rate (TPR), and True Negative Rate (TNR). Collectively, these metrics provide a quantitative assessment of the extent to which virtual staining methods correctly identify and localize target stained regions.

\textit{mIF Staining Accuracy for HEMIT~\cite{bian2024hemit}.} Because an established protocol for evaluating staining accuracy on the multiplexed immunofluorescence (mIF) HEMIT dataset is not currently available, we introduce a segmentation-based evaluation framework that extends the principles used in our IHC analysis. To assess staining accuracy in this multi-channel setting, generated mIF images are decomposed into their constituent grayscale channels corresponding to DAPI (nuclei), CD3 (T cells), and PanCK (epithelial cells). Within each channel, pixel intensity reflects the relative presence of the corresponding biomarker.

Binary masks of positive signal are extracted using Otsu thresholding applied independently to each channel on an image-by-image basis, allowing the thresholding procedure to adapt to variations in signal intensity across samples. Thresholding is followed by morphological post-processing, including dilation and median filtering, to suppress background noise and improve mask consistency. The same processing pipeline is applied to both ground-truth and generated images, with representative examples shown in Supplementary Fig.~\ref{fig:qualitative_results_threshold}.

Staining accuracy is quantified using standard segmentation metrics, including the Dice coefficient, Intersection over Union (IoU), and Hausdorff Distance (HD), computed directly between the generated and reference masks. Metrics are reported separately for each channel to capture marker-specific performance and assess both spatial overlap and boundary agreement. By complementing conventional image-fidelity metrics with channel-wise staining-accuracy measures, this protocol provides an automated and interpretable evaluation of virtual mIF generation. For reference, Table~\ref{tab:metric_table} summarizes all metrics included in our evaluation framework.

{In summary, our evaluation pipeline jointly assesses image realism and staining accuracy, providing a more complete and balanced benchmark for virtual staining models. By combining fidelity metrics with segmentation-based accuracy measures, we move beyond visual quality alone to quantify how well models recover underlying staining signals and spatial patterns. This dual perspective enables fair comparison across methods and offers a more informative assessment of their performance.}

\section{Results and Discussion}

Tables \ref{tab:results_main_table_CK818}, \ref{tab:results_main_table_CD3} and  \ref{tab:quantiative_metrics}  compare \name with paired and unpaired baselines on   CK8/18, CD3 and HEMIT dataset respectively. 
.
\begin{table*}[!htb]
\scalebox{0.82}{
\setlength{\tabcolsep}{3pt}
    \begin{tabular}{p{0.3cm}|c||c|c|c||c|c|c|c||c|c|c|c|c}
   & & \multicolumn{3}{c||}{\bf Texture Metrics } &  \multicolumn{4}{c||}{\bf Distribution Metrics} & \multicolumn{5}{c}{\bf Segmentation Metrics} \\
   & I2I Methods & PSNR $\uparrow$ & SSIM $\uparrow$ & MSE $\downarrow$& FID $\downarrow$& KID $\downarrow$ & Prec $\uparrow$ & Rec $\uparrow$  & Dice $\uparrow$ & IoU $\uparrow$ & HD$\downarrow$ & TPR$\uparrow$ & TNR$\uparrow$ \\
    \hline
    \hline
     \multirow{16}{=}{\begin{sideways}\textbf{Unpaired}\end{sideways}}   
   &    CycleGAN  \cite{zhu2017unpaired}           & 
       19.90 & 0.575 & 232.79 & 7.91 & \bf 0.0021 & 0.8707 &  0.8578 
      & 0.6236 & 0.5213 & 41.27 & 0.6786 & 0.9131 \\
   &    UNIT       \cite{liu2017unsupervised}                       & 
     12.36 & 0.000 & 296.93 & 13.04 & 0.0049 & 0.7544 & 0.6917 
     & 0.5449 & 0.4594 & 50.27 & 0.5696 & 0.8810 \\
   &    UGATIT   \cite{kim2019u}                                    & 
       19.82 & 0.586 & 233.88 & 16.27 & 0.0073 & 0.7351 & 0.7635 
       & 0.6199 & 0.5179 & 42.43 & 0.6664 & 0.9137 \\
       
   &    CUT       \cite{park2020contrastive}       & 
   19.39 & 0.560 & 237.12 & 9.34 & 0.0033 & 0.8463 & 0.8394 
   & 0.5764 & 0.4836 & 48.13 & 0.6278 & 0.9093 \\
   &    FastCUT    \cite{park2020contrastive}                       & 
      18.92 & 0.557 & 234.02 & 14.71 & 0.0088 & 0.7800 & 0.7790 
   & 0.4165 & 0.3339 & 69.79 & 0.4296 & \bf 0.9462 \\
   &    ACL GAN    \cite{zhao2020unpaired}                          & 
      14.80 & 0.50 & 295.66 & 107.33 & 0.1028 & 0.0376 & 0.0515
      & 0.4725 & 0.3539 & 47.22 & 0.7182 & 0.7109 \\
   &    NICE GAN  \cite{Chen_2020_CVPR}                             & 
      19.12 & 0.57 & 237.10 & 28.86 & 0.0186 & 0.6032 & 0.5708 
      & 0.5152 & 0.4237 & 64.83 & 0.5466 & 0.9171 \\
   &    Attn. GAN \cite{tang2021attentiongan}                   &  
    19.37 & 0.57 & 235.14 & 10.79 & 0.0040 & 0.8455 & 0.8085 
     & 0.5988 & 0.4993 & 46.31 & 0.6599 & 0.9067 \\
   &    QS GAN \cite{hu2022qs}                         & 
       19.41 & 0.56 & 236.87 & 8.91 & 0.0024 & 0.8497 & 0.8254 
       & 0.5658 & 0.4749 & 48.29 & 0.6090 & 0.9078 \\
   &    Decent        \cite{xieunsupervised}                        & 
    19.64 & 0.55 & 239.42 & 10.44 & 0.0042 & 0.8349 & 0.8363 
& 0.6422 & 0.5411 & 40.41 & 0.7016 & 0.9084 \\
   &    VQ-I2I-Un       \cite{chen2022eccv}                   & 
       12.56 & 0.13 & 303.43 & 24.08 & 0.0116 & 0.4988 & 0.6904 
       & 0.2505 & 0.1690 & 78.41 & 0.2665 & 0.7414 \\
   &    UVCGAN    \cite{torbunov2023uvcgan}                         & 
       19.30 & 0.57 & 237.25 & 19.69 & 0.0118 & 0.7131 & 0.7578
  & 0.5814 & 0.4808 & 49.48 & 0.6520 & 0.8845 \\
   &    SANTA     \cite{xie2023unpaired}                            & 
   18.98 & 0.52 & 247.15 & 11.41 & 0.0043 & 0.8175 & 0.7747
  & 0.6007 & 0.4977 & 43.84 & 0.6508 & 0.8991 \\
   &    UNSB      \cite{kim2023unpaired}                            & 
       18.88 & 0.54 & 239.71 & 13.20 & 0.0060 & 0.8306 & 0.5904 
   & 0.6190 & 0.5225 & 49.66 & 0.6377 & 0.9162 \\
   &    StegoGAN \cite{wu2024stegogan}                              & 
       19.41 & 0.57 & 234.33 & 9.93 & 0.0033 & 0.8603 & 0.8140 
     & 0.6237 & 0.5199 & 42.35 & 0.6937 & 0.9036 \\  
       \hline
       \hline
    \multirow{9}{=}{\begin{sideways}\textbf{Paired}\end{sideways}}   
   &    Pix2Pix    \cite{isola2017image}           & 
       18.78 & 0.48 & 250.81 & 44.53 & 0.0241 & 0.7368 & 0.7158 
      & 0.5824 & 0.4949 & 53.29 & 0.6111 & 0.9289 \\
   &    PyramidPix2Pix \cite{liu2022bci}           & 
       21.37 & 0.60 & 224.87 & 28.74 & 0.0220 & 0.8172 & 0.7030  
      & 0.6887 & 0.5937 & 32.50 & 0.7021 &   {0.9411} \\
 & AdaptiveNCE \cite{li2023adaptive} &   {21.38} &   {0.61} &   {222.95} & \bf 7.33 &   {0.0022} &   {0.9034} & \bf 0.9237
 &   {0.6898} &   {0.5941} &   {32.11} & \bf 0.7358 & 0.9358 \\
   &    VQ-I2I-P       \cite{chen2022eccv}                     & 
     12.54 & 0.13 & 305.22 & 29.26 & 0.0197 & 0.5013 & 0.6133 
      & 0.1472 & 0.0927 & 105.39 & 0.1572 & 0.8405 \\
   &    {EDSDE}       \cite{zhao2022egsde}                            & 
   14.83 & 0.27 & 300.17 & 92.88 & 0.0841 & 0.1654 & 0.5075 
 & -- & -- & -- & -- & -- \\
  &    {CycleDiffusion} \cite{wu2022unifying}                        & 
 15.73 & 0.44 & 300.17 & -- & -- & -- & -- & 0.3843 & 0.2759 & 74.22 & 0.5589 & 0.7824 \\
   &    {BBDM}    \cite{li2023bbdm}                                   & 
       20.03 & 0.57 & 224.63 & 17.41 & 0.0096 & 0.8054 & 0.7842 
       & 0.6315 & 0.5329 & 40.32 & 0.6284 & 0.9396 \\
   &   {LBBDM-F4}    \cite{li2023bbdm}                         & 
       18.45 & 0.45 & 257.44 & 18.32 & 0.0119 & 0.7304 & 0.7573 
       & 0.5579 & 0.4748 & 52.73 & 0.5727 & 0.9245 \\
   &    {LBBDM-F16}    \cite{li2023bbdm}                        & 
       16.20 & 0.26 & 277.92 & 61.16 & 0.0513 & 0.6102 & 0.1671 
       & 0.5005 & 0.4195 & 63.81 & 0.5294 & 0.8974 \\
 & DualBranch \cite{bian2024hemit}& 21.00 & 0.5609 & 226.78 & 14.23 & 0.0086 & 0.8974 &0.8778 & 0.6903 & 0.5973 & 32.65 &  0.7244 & 0.9301 \\
 & PPT \cite{zhang2024high} & 20.26  & 0.5415 & 241.9 & 22.11 &  0.0156 & 0.7808 & 0.7101 & 0.6268
 & 0.5382 & 41.23 & 0.6454 & 0.9256 \\
  & MCS-Stain \cite{hu2025mcs} & 20.94 & 0.5604 & 228.49 & 10.07  & 0.0049 & 0.9086 &  0.9050 &
0.6804 & 0.5862 & 31.70 & 0.6923 &  0.9424 \\
  & M2PL-GAN \cite{liu2026m2pl} & 20.52  & 0.5778 & 231.46 & 10.04 & 0.0037 & 0.8733 & 0.8456 &
 0.6716 & 0.5751 & 34.74 & 0.7127 & 0.9228  \\
     
     &   \name (\textbf{Ours})            &  \bf 22.24$^*$	& \bf 0.644$^*$	& \bf 219.19$^*$ & 17.69 & 0.0107 &	\bf 0.9131 &	  {0.8410} &\bf 0.7049$^*$	& \bf 0.6149$^*$	& \bf 30.19$^*$	&   {0.7298}$^*$	& 0.9351 \\
    \end{tabular}}

 \caption{\textbf{Quantitative Results on the CK8/18 Virtual IHC Staining Dataset.} \textbf{Bold} indicates the best performance for each metric. Across conventional image-fidelity metrics, \name achieves the best PSNR, SSIM, and MSE. More importantly, on segmentation-based metrics that directly assess staining accuracy, \name obtains the highest Dice and IoU scores and the lowest Hausdorff Distance, indicating improved recovery of the spatial distribution and boundaries of CK8/18-positive regions. These results demonstrate that \name performs competitively across a broad range of evaluation criteria while achieving the strongest overall performance on the primary staining-accuracy metrics. This table only states the Mean values of Texture and Segmentation Metrics, for mean with their standard deviations refer to Supplementary Table \ref{tab:metrics_ck818_mean_stddev}. As AdaptiveNCE \cite{li2023adaptive} was the strongest baseline among the compared methods, we performed paired statistical significance tests to compare its performance with our method on texture and segmentation metrics. Metrics marked with an asterisk (*) indicate that the observed improvement is statistically significant ($p < 0.001$).}
    \label{tab:results_main_table_CK818}
\end{table*}

\textbf{\name consistently achieves good performance on image-fidelity metrics across multiple virtual staining tasks.} The results in Tables~\ref{tab:results_main_table_CK818}, \ref{tab:results_main_table_CD3}, and \ref{tab:quantiative_metrics} show that the proposed model achieves the highest PSNR and lowest MSE across all evaluated datasets. It also performs competitively in terms of structural similarity, attaining the highest SSIM on the CK8/18 and CD3 datasets and ranking third on the HEMIT dataset, where MCS-Stain achieves the best score. Collectively, these results indicate improved reconstruction quality, characterized by reduced pixel-wise error and strong preservation of structural information relative to both paired and unpaired baselines. Since PSNR, SSIM, and MSE are standard measures of image fidelity, the results suggest that \name generates virtual stains that closely resemble the corresponding ground-truth images in both appearance and structure.

Among paired methods, AdaptiveNCE, PyramidPix2Pix and MCS-Stain emerge as the strongest competitors, ranking among the top-performing approaches across multiple datasets. For unpaired methods, performance varies depending on the staining task. CycleGAN and UGATIT achieve strong results on CK8/18, QS-GAN performs best on the HEMIT dataset, and CycleGAN and FastCUT rank among the strongest unpaired methods on the CD3 dataset. These observations suggest that the relative strengths of unpaired approaches are task dependent. Across all three datasets, \name consistently outperforms other convolution--transformer-based (hybrid backbone) image translation models(DualBranch Pix2Pix and UVCGAN). This trend suggests that the proposed neural implicit formulation provides advantages beyond architectural improvements alone. Overall, the results demonstrate that \name maintains competitive and consistent performance across diverse staining tasks, datasets, and evaluation settings.
\begin{table*}[!htb]
\scalebox{0.82}{
\setlength{\tabcolsep}{3pt}
\centering
\begin{tabular}{p{0.3cm}|c||c|c|c||c|c|c|c||c|c|c|c|c}
 & & \multicolumn{3}{c||}{\bf Texture Metrics} & \multicolumn{4}{c||}{\bf Distribution Metrics} & \multicolumn{5}{c}{\bf Segmentation Metrics} \\
 & I2I Methods & PSNR $\uparrow$ & SSIM $\uparrow$ & MSE $\downarrow$ & FID $\downarrow$ & KID $\downarrow$ & Prec $\uparrow$ & Rec $\uparrow$ & Dice $\uparrow$ & IoU $\uparrow$ & HD$\downarrow$ & TPR$\uparrow$ & TNR$\uparrow$ \\
\hline
\hline
\multirow{16}{=}{\begin{sideways}\textbf{Unpaired}\end{sideways}}
& CycleGAN~\cite{zhu2017unpaired} &
19.42 & 0.5477 & 219.61 & 6.82 & 0.0021 & 0.8621 & 0.8322 &
0.5432 & 0.3981 & 21.51 & 0.5362 & 0.9858 \\
& UNIT~\cite{liu2017unsupervised} &
17.28 & 0.4959 & 249.07 & 10.89 & 0.0053 & 0.8319 & 0.7822 &
0.4556 & 0.3178 & 26.45 & 0.4644 & 0.9823 \\
& UGATIT~\cite{kim2019u} &
19.21 & 0.5538 & 218.57 & 9.22 & 0.0043 & 0.8365 & 0.7779 &
0.4894 & 0.3484 & 28.46 & 0.4533 & 0.9888 \\
& CUT~\cite{park2020contrastive} &
19.49 & 0.5377 & 221.37 & \textbf{5.95} & 0.0014 & 0.8839 & \textbf{0.8345} &
0.5117 & 0.3691 & 23.03 & 0.5027 & 0.9850 \\
& FastCUT~\cite{park2020contrastive} &
19.50 & 0.5383 & 221.46 & 13.51 & 0.0080 & 0.8699 & 0.7351 &
0.5026 & 0.3587 & 23.53 & 0.4807 & 0.9860 \\
& ACL GAN~\cite{zhao2020unpaired} &
16.10 & 0.4650 & 262.51 & 14.96 & 0.0087 & 0.7143 & 0.8071 &
0.3638 & 0.2326 & 27.63 & 0.3798 & 0.9664 \\
& NICE GAN~\cite{Chen_2020_CVPR} &
18.94 & 0.5469 & 226.43 & 21.35 & 0.0163 & 0.7034 & 0.7345 &
0.4925 & 0.3516 & 24.52 & 0.4763 & 0.9866 \\
& Attn. GAN~\cite{tang2021attentiongan} &
19.36 & 0.5474 & 219.33 & 7.16 & 0.0019 & 0.8589 & 0.8341 &
0.5352 & 0.3908 & 22.06 & 0.5323 & 0.9856 \\
& QS GAN~\cite{hu2022qs} &
19.39 & 0.5378 & 222.11 & 6.19 & \textbf{0.0013} & 0.8744 & 0.8286 &
0.5027 & 0.3597 & 22.98 & 0.4906 & 0.9849 \\
& Decent~\cite{xieunsupervised} &
19.20 & 0.5241 & 226.98 & 6.79 & 0.0016 & 0.8602 & 0.8052 &
0.5049 & 0.3583 & 23.15 & 0.5107 & 0.9828 \\
& VQ-I2I-Un~\cite{chen2022eccv} &
18.22 & 0.3407 & 271.45 & 22.68 & 0.0133 & 0.7086 & 0.4973 &
0.3885 & 0.2561 & 33.40 & 0.3999 & 0.9780 \\
& UVCGAN~\cite{torbunov2023uvcgan} &
19.47 & 0.5452 & 222.93 & 11.33 & 0.0054 & 0.8072 & 0.7258 &
0.5345 & 0.3901 & 21.74 & 0.5569 & 0.9826 \\
& SANTA~\cite{xie2023unpaired} &
18.90 & 0.5168 & 227.70 & 7.45 & 0.0030 & 0.8407 & 0.8205 &
0.4628 & 0.3153 & 22.25 & 0.4840 & 0.9716 \\
& UNSB~\cite{kim2023unpaired} &
19.31 & 0.5276 & 225.73 & 17.22 & 0.0128 & 0.6375 & 0.8037 &
0.4877 & 0.3460 & 23.95 & 0.4738 & 0.9863 \\
& StegoGAN~\cite{wu2024stegogan} &
19.38 & 0.5468 & 220.74 & 7.08 & 0.0020 & 0.8613 & 0.8326 &
0.5311 & 0.3865 & 22.37 & 0.5178 & 0.9863 \\
\hline
\hline
\multirow{9}{=}{\begin{sideways}\textbf{Paired}\end{sideways}}
& Pix2Pix~\cite{isola2017image} &
19.01 & 0.4628 & 234.63 & 19.19 & 0.0089 & 0.7962 & 0.6342 &
0.5427 & 0.3947 & 23.86 & 0.5596 & 0.9825 \\
& PyramidPix2Pix~\cite{liu2022bci} &
20.09 & 0.5443 & 215.69 & 28.03 & 0.0251 & 0.6991 & 0.7494 &
0.5935 & 0.4461 & 18.59 & 0.5978 & 0.9855 \\
& AdaptiveNCE~\cite{li2023adaptive} &
20.45 & 0.5667 & 213.23 & 11.22 & 0.0045 & 0.8758 & 0.8195 &
0.5932 & 0.4482 & \textbf{17.58} & 0.5987 & 0.9851 \\
& VQ-I2I-P~\cite{chen2022eccv} &
16.39 & 0.2986 & 253.03 & 110.30 & 0.0774 & 0.5366 & 0.1217 &
0.2896 & 0.1852 & 70.38 & 0.2546 & \textbf{0.9931} \\
& EDSDE~\cite{zhao2022egsde} &
14.54 & 0.2630 & 262.56 & 104.50 & 0.0555 & 0.5281 & 0.2058 &
- & - & - & - & - \\
& CycleDiffusion~\cite{wu2022unifying} &
15.31 & 0.4092 & 312.48 & 72.28 & 0.0560 & 0.7610 & 0.3980 &
0.3974 & 0.2617 & 31.69 & 0.5083 & 0.9605 \\
& BBDM~\cite{li2023bbdm} &
19.80 & 0.5468 & 212.80 & 21.56 & 0.0162 & 0.7531 & 0.6820 &
0.5835 & 0.4372 & 19.12 & 0.6140 & 0.9824 \\
& LBBDM-F4~\cite{li2023bbdm} &
18.46 & 0.4391 & 244.84 & 16.20 & 0.0105 & 0.7776 & 0.6918 &
0.3900 & 0.2550 & 32.83 & 0.3836 & 0.9799 \\
& LBBDM-F16~\cite{li2023bbdm} &
16.94 & 0.2935 & 266.40 & 64.77 & 0.0554 & 0.1330 & 0.5286 &
0.2581 & 0.1537 & 47.14 & 0.2428 & 0.9809 \\
& DualBranch~\cite{bian2024hemit} &
20.21 & 0.5167 & 219.88 & 10.55 & 0.0061 & 0.9006 & 0.7187 &
0.5966 & 0.4382 & 19.75 & 0.5519 & 0.9900 \\
& PPT \cite{zhang2024high} & 19.83 & 0.4915 & 230.51 & 16.68 & 0.0125 & 0.8718 & 0.5093 &
 0.3325 & 0.2223 & 27.23 & 0.3409 & 0.9207 \\
& MCS-Stain \cite{hu2025mcs} & 20.01  & 0.506 & 223.90 & 10.39 & 0.0068 & \bf 0.9111 & 0.7586 & 0.5980
& 0.4516  & 17.58 & \bf 0.6120  &  0.9844 \\
& M2PL-GAN \cite{liu2026m2pl} & 20.28 & 0.5562 & 216.70 & 8.69 & 0.0038 &  0.8761& 0.8250 & 0.5737 
 & 0.4264 & 18.59 & 0.5809 & 0.9863 \\
& \name (\textbf{Ours}) &
\textbf{21.28}$^*$ & \textbf{0.5937}$^*$ & \textbf{211.91}$^*$ & 33.79 & 0.0334 & 0.7622 & 0.6709 &
\textbf{0.6121}$^*$ & \textbf{0.4617}$^*$ & 17.77 & 0.5980$^*$ & 0.9877 \\
\end{tabular}}
\caption{\textbf{Quantitative Results for the CD3 Dataset.} \textbf{Bold} indicates the best performance for each metric. Across conventional image-fidelity metrics, \name achieves the best PSNR, SSIM, and MSE. More importantly, on segmentation-based metrics that directly assess staining accuracy, \name obtains the highest Dice and IoU scores while remaining competitive on the remaining metrics. This table only states the Mean values of Texture and Segmentation Metrics, for mean with their standard deviations refer to Supplementary Table \ref{tab:metrics_cd3_mean_stddev}. As AdaptiveNCE \cite{li2023adaptive} and MCS-Stain\cite{hu2025mcs} was the strongest baseline among the compared methods, we performed paired statistical significance tests to compare its performance with our method on texture and segmentation metrics. Metrics marked with an asterisk (*) indicate that the observed improvement is statistically significant ($p < 0.001$).}
\label{tab:results_main_table_CD3}
\end{table*}

\textbf{\name balances image realism with spatial accuracy.} While some unpaired methods, such as CUT and QS-GAN, achieve strong performance on distribution-based metrics (e.g., FID), these improvements do not necessarily translate to accurate localization of staining patterns. For example, on the HEMIT dataset (Table~\ref{tab:quantiative_metrics}), CUT achieves a low FID score of 8.05 but attains substantially lower segmentation accuracy on the CD3 channel (Dice: 0.0276). This observation highlights a limitation of relying solely on distribution-based metrics: visually realistic images may not faithfully recover the spatial distribution of target staining patterns. In contrast, \name is trained using paired supervision and optimized for direct pixel-level correspondence between source and target domains. As reflected by the segmentation metrics, this formulation enables \name to better preserve the spatial localization of staining signals while maintaining competitive image quality.

\textbf{\name achieves strong staining accuracy, particularly on challenging sparse markers.} Tables~\ref{tab:results_main_table_CK818}, \ref{tab:results_main_table_CD3}, and \ref{tab:quantiative_metrics} show that paired methods generally outperform unpaired methods on segmentation-based metrics, highlighting the value of paired supervision for accurate spatial localization. Across all evaluated datasets, \name achieves the highest Dice and IoU scores while remaining highly competitive in terms of boundary accuracy, as measured by Hausdorff Distance.

The results also highlight the varying difficulty of different virtual staining tasks. Translating broadly distributed markers such as DAPI and PanCK is comparatively easier because the target signal occupies larger spatial regions. In contrast, sparse markers such as CD3 require accurate localization of relatively rare staining patterns, making the task substantially more challenging. Under these conditions, \name exhibits particularly strong performance. For example, on the HEMIT CD3 channel (Table~\ref{tab:quantiative_metrics}), \name achieves a Dice score of 0.5026, compared with 0.3648 for the next-best baseline (MCS-Stain). Similar trends are observed across the remaining HEMIT channels, where \name consistently attains the strongest segmentation performance.

Furthermore, although some methods occasionally achieve higher True Negative Rates (TNR), these gains are often accompanied by reduced sensitivity, suggesting a tendency toward conservative predictions. In contrast, \name maintains a balanced precision--recall and TPR--TNR profile while achieving the highest overall overlap-based metrics. Taken together, these results indicate that \name more accurately recovers the spatial distribution of target staining patterns, particularly in challenging settings involving sparse or highly localized markers.

 \begin{table*}[!thb]
\setlength{\tabcolsep}{2pt}
\scalebox{0.76}{
    \begin{tabular}{p{0.25cm}|c||ccc||ccc||ccc|ccc|ccc}
   & & \multicolumn{3}{c||}{\bf Texture Metrics } &  \multicolumn{3}{c||}{\bf Distribution Metrics} & \multicolumn{9}{c}{\bf Segmentation Metrics}\\ 
   & & \multicolumn{3}{c||}{ } &  \multicolumn{3}{c||}{} & \multicolumn{3}{c|}{\it DAPI} & \multicolumn{3}{c|}{\it CD3} &  \multicolumn{3}{c}{\it PanCK}\\  
   & Methods & PSNR $\uparrow$ & SSIM $\uparrow$ & MSE$\downarrow$& FID $\downarrow$ & Prec $\uparrow$ & Recall $\uparrow$&  Dice $\uparrow$ & IoU $\uparrow$ & HD $\downarrow$ &  Dice $\uparrow$ & IoU $\uparrow$ & HD $\downarrow$ &  Dice $\uparrow$ & IoU $\uparrow$ & HD $\downarrow$\\
    \hline
    \hline
    \multirow{11}{=}{\begin{sideways}\textbf{Unpaired Models}\end{sideways}} 
         & CycleGAN	\cite{zhu2017unpaired}          &  21.32 & 0.5455 & 149.93 & 36.80 & 0.7106 & 0.2926 & 0.2798 & 0.2007 & 101.51 & 0.0317 & 0.0169 & 104.85 & 0.0465 & 0.0248 & 64.56 \\
          & UNIT	\cite{hu2021unit}                   &  19.54 & 0.6454 & 136.39 & 12.24 & 0.7177 & 0.5606 & 0.4498 & 0.3687 & 61.55 & 0.1829 & 0.1225 & 108.58 & 0.6820 & 0.5754 & 25.84 \\
           & UGATIT \cite{kim2019u}	                & 25.19 & 0.7423 & 125.97 & 19.58 & 0.7201 & 0.5103 & 0.4436 & 0.3583 & 68.97 & 0.0200 & 0.0106 & 108.86 & 0.5424 & 0.4283 & 33.14 \\
         & CUT	   \cite{park2020contrastive}       & 21.37 & 0.5055 & 165.37 &   {8.05} & 0.7660 & 0.6501 & 0.0772 & 0.0483 & 110.81 & 0.0276 & 0.0149 & 108.28 & 0.0708 & 0.0391 & 60.16 \\
         & FastCUT	\cite{park2020contrastive}      & 26.01 & 0.7541 & 94.44 & 15.85 & 0.8086 & 0.6513 & 0.5125 & 0.4142 & 64.18 & 0.1899 & 0.1278 & 88.21 & 0.7709 & 0.6587 & 18.03 \\
         & NICE GAN  \cite{Chen_2020_CVPR}           &  24.36 & 0.54 &  112.19 & 503.78 & 0.0000 & 0.0000& 0.0000 & 0.0000& 0.0000 & 0.0000& 0.0000 & 0.0000& 0.0000 & 0.0000& 0.0000 \\
         & AttentionGAN	\cite{tang2021attentiongan} &  25.45 & 0.7604 & 91.15 & 12.88 & 0.7841 & 0.5683 & 0.4858 & 0.3894 & 65.80 & 0.0125 & 0.0067 & 100.37 & 0.7179 & 0.5921 & 26.06 \\
         & QS-GAN	\cite{hu2022qs}                 & 27.31 & 0.8164 & 73.76 & \bf 6.90 & 0.8478 & 0.7555 & 0.5367 & 0.4440 & 50.62 & 0.2625 & 0.1823 & 79.72 & 0.8089 & 0.7087 & 13.70 \\
         & DECENT \cite{xieunsupervised}             &  23.44 & 0.6527 & 124.44 & 8.24 & 0.7733 & 0.6305 & 0.2009 & 0.1304 & 98.06 & 0.1615 & 0.1057 & 93.46 & 0.5139 & 0.3805 & 41.44 \\
         & VQ-I2I-UnParied \cite{chen2022eccv}        & 22.88 & 0.5784 & 125.32 & 61.00 & 0.4294 & 0.0517 & 0.1942 & 0.1206 & 95.64 & 0.0642 & 0.0371 & 104.59 & 0.2568 & 0.1543 & 55.67 \\
         & UVCGAN \cite{torbunov2023uvcgan}	        & 23.26 & 0.6215 & 96.23 & 69.99 & 0.5748 & 0.1691 & 0.0716 & 0.0411 & 123.00 & 0.0053 & 0.0028 & 132.99 & 0.0235 & 0.0127 & 65.45 \\
         & UNSB      \cite{kim2023unpaired}          &  21.90 & 0.5350 & 157.75 & 13.99 & 0.7515 & 0.6294 & 0.1719 & 0.1132 & 103.49 & 0.0311 & 0.0169 & 120.74 & 0.2226 & 0.1343 & 58.74 \\
           & StegoGAN	\cite{wu2024stegogan}           &  22.11 & 0.5894 & 134.48 & 46.90 & 0.5748 & 0.1691 & 0.2969 & 0.2174 & 101.48 & 0.0248 & 0.0132 & 108.23 & 0.0236 & 0.0126 & 80.50 \\
    
         \hline
         \hline
    \multirow{7}{=}{\begin{sideways}\textbf{Paired }\end{sideways}}   
        & Pix2Pix \cite{isola2017image}	     & 28.61 & 0.8372 & 64.01 & 22.82 & 0.7068 & 0.4861 & 0.5428 & 0.4518 & 52.93 & 0.2419 & 0.1660 & 73.68 & 0.8042 & 0.7127 & 13.84 \\
        & PyramidPix2Pix \cite{liu2022bci}   &   {29.26}&   0.8522 &   {61.28} & 10.51 &   {0.8290} & 0.7268 & 0.5686 & 0.4769 & 49.72 &   {0.3157} &   {0.2318} &   {68.23} &   {0.8206} &   {0.7274} &   {12.92} \\
         & AdaptiveNCE	\cite{li2023adaptive}       &  27.11 & 0.4524 & 129.83 & 165.55 & 0.3198 & 0.0032 &   {0.6063} &   {0.5332} &   {48.64} & 0.0965 & 0.0558 & 110.95 & 0.8114 & 0.7013 & 13.72 \\
        & VQ-I2I-Paried \cite{chen2022eccv}	 &  26.04 & 0.7787 & 81.53 & 26.12 & 0.6548 & 0.6548 & 0.5102 & 0.4145 & 62.90 & 0.2566 & 0.1720 & 88.45 & 0.7441 & 0.6125 & 17.43 \\
       
          & DualBranch \cite{bian2024hemit}& 28.07 & 0.8209  &  68.322 & 42.59 & 0.6644 & 0.2557 & 0.5164 & 0.4217 & 57.70 & 0.2073 &  0.1422 & 80.61 & 0.7802 & 0.6830 &  17.41 \\
          & PPT \cite{zhang2024high} & 27.05 &  0.801 & 77.79 & 16.46 & 0.7972 & 0.4661 & 0.4992 & 0.3357 & 60.33 & 0.2139 &  0.1403 & 85.04 & 0.7716 & 0.6536 & 15.74 \\    
          & MCS-Stain \cite{hu2025mcs} & 30.87 & \bf 0.8763 & 55.82& 9.68 & 0.8581 & \bf 0.7787 & 0.6090 & 0.5267 & 49.87 & 0.3648 &  0.2839 & 67.63 & 0.8370 & 0.7511 & 13.86 \\    
        & DiffVS* \cite{oh2026virtual} & 30.60 & 0.836 & - & - & - & - & - & - & - & - &  - & - & - & - & - \\
         & M2PL-GAN \cite{liu2026m2pl} & 28.36 & 0.8404 & 63.71 & 9.03 & 0.8607 & 0.6793 & 0.5641 & 0.4745 & 52.41 & 0.2673 &  0.1839 & 78.57 & 0.8285 & 0.7354 & 12.83 \\    

         &  \name (\textbf{Ours})    &  \bf 32.26$^*$ &  0.8450$^*$ & \bf 55.38$^*$ &  12.01$^*$ & \bf 0.8685$^*$ &  0.7776$^*$ & \bf 0.6355$^*$ & \bf 0.5496$^*$ & \bf 43.78$^*$ & \bf 0.5026$^*$ & \bf 0.4125$^*$ & \bf 58.30$^*$& \bf 0.8538$^*$ & \bf 0.7713$^*$ & \bf 12.58$^*$ \\
    \end{tabular}}
\caption{\textbf{Quantitative Results on the HEMIT Dataset.} \textbf{Bold} indicates the best performance for each metric. While competing methods achieve the best performance on isolated image-fidelity metrics such as SSIM and FID, \name attains the strongest overall performance on segmentation-based staining-accuracy metrics across all three mIF channels (DAPI, CD3, and PanCK). These results indicate that \name more accurately recovers the spatial distribution and boundaries of target staining patterns. This table only states the Mean values of Texture and Segmentation Metrics, for mean with their standard deviations refer to Supplementary Table \ref{tab:hemit_metrics_mean_std}.  MCS-Stain\cite{hu2025mcs} was the strongest baseline among the compared methods, we performed paired statistical significance tests to compare its performance with our method on texture and segmentation metrics. Metrics marked with an asterisk (*) indicate that the observed improvement is statistically significant ($p < 0.001$).
}
    \label{tab:quantiative_metrics}
\end{table*}

Among the paired approaches, AdaptiveNCE, PyramidPix2Pix, and MCS-Stainer again emerge as the strongest competitors in terms of staining accuracy. In contrast, the best-performing unpaired method varies across tasks, with UGATIT achieving the highest segmentation accuracy on CK8/18, CycleGAN on CD3, and QS-GAN on HEMIT. Overall, paired methods generally outperform unpaired approaches in segmentation-based evaluations, highlighting the importance of direct supervision for preserving stain-specific spatial information. This finding is consistent with prior studies \cite{dubey2023structural,kataria2025building,li2023adaptive}, which reported a weak correspondence between image-fidelity metrics and staining accuracy. Because unpaired methods are optimized to align source and target distributions without enforcing pixel-level correspondence, they can achieve competitive distribution-based scores while exhibiting inferior localization performance. Consequently, our results further emphasize the importance of incorporating staining-accuracy metrics alongside FID, as favorable distributional similarity does not necessarily translate to the accurate reconstruction of biologically meaningful spatial patterns.

\textbf{Challenges with Diffusion- and VQ-Based Architectures.} While diffusion models and vector-quantized frameworks have demonstrated strong performance in natural image generation, our results suggest that these approaches can be less effective for virtual staining tasks that require precise spatial correspondence between source and target domains. Across Tables~\ref{tab:results_main_table_CK818} and \ref{tab:results_main_table_CD3}, methods such as EDSDE, CycleDiffusion, and VQ-I2I (both paired and unpaired variants) generally achieve lower image-fidelity and segmentation-based metrics than the strongest GAN-based baselines and \name. For example, on the CK8/18 dataset (Table~\ref{tab:results_main_table_CK818}), EDSDE and VQ-I2I-P achieve PSNR values of 14.83 and 12.54, respectively, substantially below those of the leading methods. These findings suggest that preserving fine-grained spatial correspondence remains challenging for certain diffusion- and VQ-based image translation frameworks in the virtual staining setting.

\textbf{\name exhibits strong cross-dataset consistency.} An important characteristic of \name is its consistent performance across datasets with differing staining targets and image characteristics. While several baseline methods perform strongly on individual datasets, their relative ranking often varies across tasks. For example, AdaptiveNCE is among the strongest competing methods on the CK8/18 and CD3 datasets (Tables~\ref{tab:results_main_table_CK818} and \ref{tab:results_main_table_CD3}), achieving competitive image-fidelity and segmentation scores. However, on the HEMIT dataset (Table~\ref{tab:quantiative_metrics}), its performance declines substantially, particularly on several segmentation-based metrics and distribution-based measures such as FID. In contrast, \name maintains competitive performance across all three datasets and consistently achieves the strongest overall staining-accuracy metrics. These results suggest that the proposed neural implicit formulation generalizes effectively across diverse virtual staining tasks and is less sensitive to variations in staining targets and dataset characteristics.

\begin{figure*}[!htb]
    \centering
    \includegraphics[trim={0.0cm 0.5cm 10.8cm 0.0cm}, clip=true, width=1.0\linewidth]{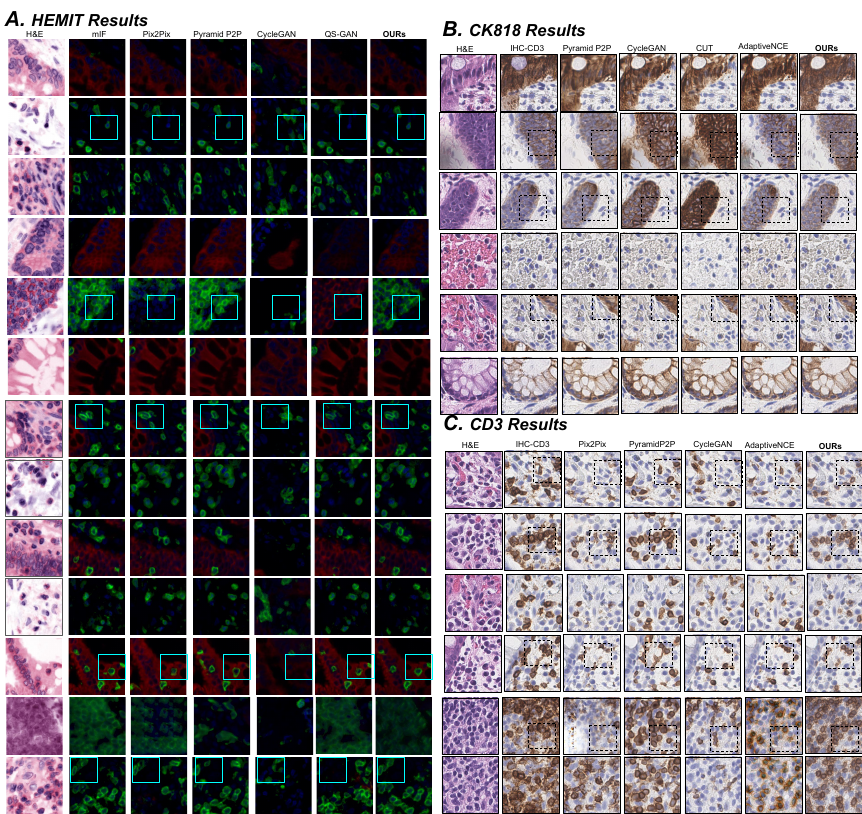}
     \caption{\textbf{Qualitative comparison of virtual staining methods.} (A) Results on the HEMIT dataset. (B) Results on the CK8/18 virtual IHC staining task. (C) Results on the CD3 virtual IHC staining task. For clarity, only the strongest paired and unpaired baselines are shown. Highlighted regions indicate representative areas with noticeable differences in stain localization, structural detail, and staining intensity. }
     \label{fig:qualitative_results}
\end{figure*}

\textbf{Qualitative results suggest improved localization and stain specificity with \name.} Representative qualitative results are shown in Fig.~\ref{fig:qualitative_results}A--C. On the HEMIT dataset, \name produces staining patterns that are visually well aligned with the reference images and exhibits fewer apparent localization errors than many competing methods. CycleGAN often generates visually plausible outputs but occasionally introduces staining patterns that are not well aligned with the ground-truth channels. PyramidPix2Pix is the strongest visual competitor; however, several examples (e.g., rows 3, 5, 7, and 15) show discrepancies between the generated and reference staining patterns. Overall, \name produces consistent staining predictions across diverse tissue morphologies and staining targets.

For CK8/18, both AdaptiveNCE and \name generate outputs whose color distributions closely resemble the reference stains, whereas PyramidPix2Pix tends to produce smoother outputs with reduced edge definition. Several competing methods also appear to overemphasize CK8/18-positive regions, resulting in darker and less nuanced staining patterns. More broadly, CK8/18 appears to be a comparatively straightforward staining task, as most methods correctly identify a large proportion of positive regions. Consequently, differences between methods are most apparent at the cellular level, including the preservation of cell boundaries and local variations in staining intensity. In these aspects, \name more closely reflects the reference images and maintains clearer delineation of individual structures.

The CD3 task is considerably more challenging. Unpaired methods such as CycleGAN~\cite{zhu2017unpaired} generally capture the overall color distribution but often exhibit less accurate localization of CD3-positive cells, particularly in rows 1, 2, 5, and 6. Paired approaches, including Pix2Pix~\cite{isola2017image}, PyramidPix2Pix~\cite{liu2022bci}, AdaptiveNCE~\cite{li2023adaptive}, and \name, demonstrate improved correspondence with the reference stains. Among these methods, \name frequently produces more localized staining patterns that align closely with individual positive cells. AdaptiveNCE occasionally introduces additional stained regions not observed in the reference images (e.g., rows 5 and 6), whereas PyramidPix2Pix exhibits reduced localization accuracy in several of the later examples. Although all methods show some degree of error on this challenging task, \name generally demonstrates stronger agreement with the reference staining patterns and fewer apparent localization errors. Additional qualitative results for CD3 and CK8/18 are provided in Supplementary Figures~\ref{fig:qualitative_results_add_cd3} and~\ref{fig:qualitative_results_add_Ck818}, respectively.

\subsection{Correlation Analysis} 


\textbf{Correlation Analysis of Evaluation Metrics on the HEMIT Dataset.} To better understand the relationships among different evaluation metrics and their implications for virtual staining performance, we conduct a correlation analysis across all reported metrics for different models on HEMIT dataset. Figure~\ref{fig:correlation_hemit} presents the lower-triangular correlation matrix computed using the average performance of all evaluated methods. The analysis includes pixel-level metrics, distribution-based metrics, and segmentation-based staining-accuracy metrics.

\begin{figure}[!thb]
    \centering
    \includegraphics[trim={0cm 0.3cm 0.2cm 1.8cm}, clip=true, width=1.0\linewidth]{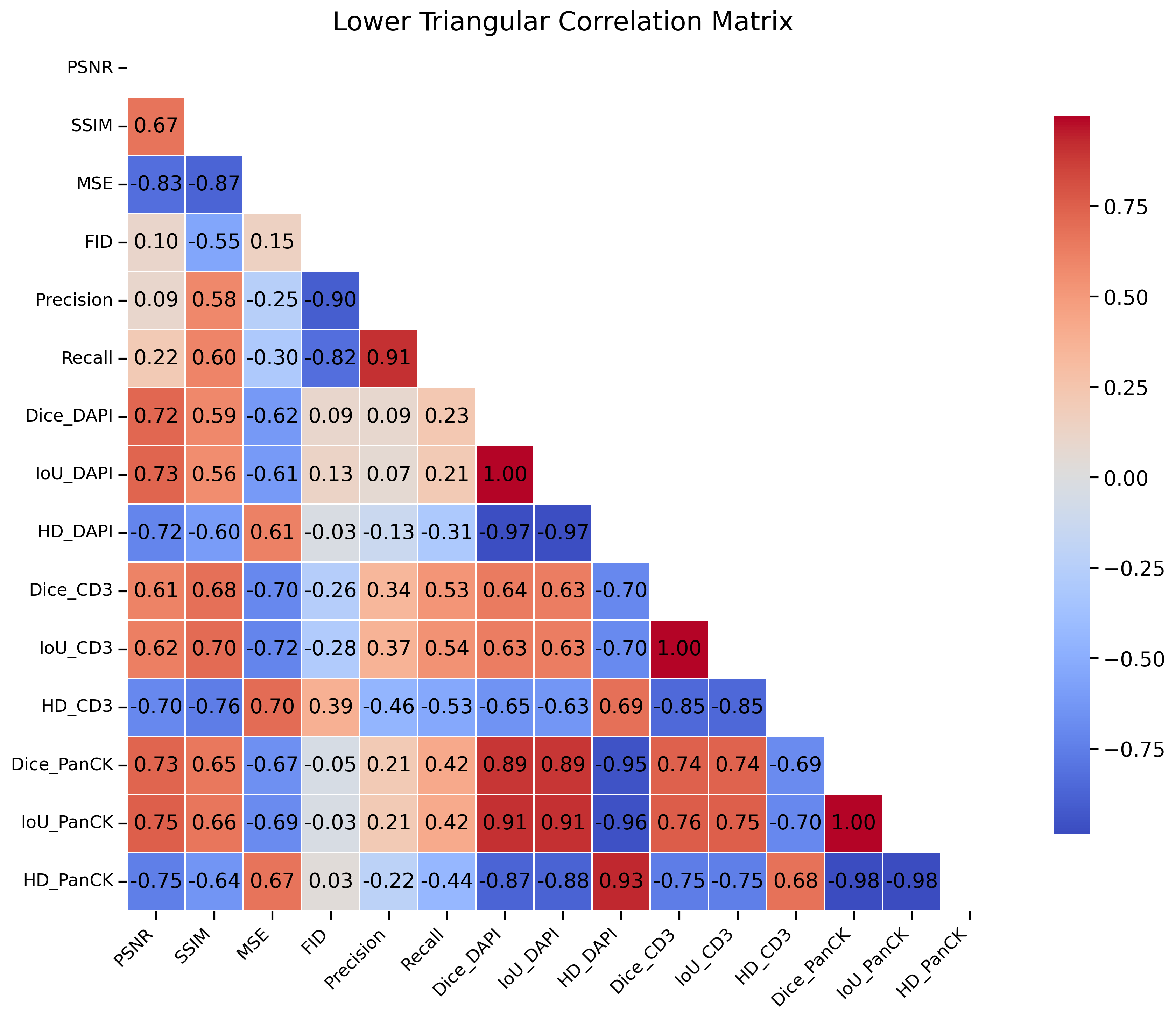}
  \caption{\textbf{Correlation analysis of evaluation metrics on the HEMIT dataset.} Lower-triangular correlation matrix showing relationships among image-fidelity, distribution-based, and staining-accuracy metrics across all evaluated methods. Strong correlations are observed within metric families (e.g., PSNR--SSIM and Dice--IoU), whereas FID exhibits relatively weak correlations with segmentation-based measures. This separation indicates that image realism and staining accuracy capture distinct aspects of virtual staining performance and should therefore be evaluated jointly rather than in isolation.
}
    \label{fig:correlation_hemit}
\end{figure}

We observe a positive correlation between PSNR and SSIM ($r > 0.65$), indicating that both metrics capture related aspects of image fidelity and reconstruction quality. In contrast, MSE shows a strong negative correlation with both PSNR and SSIM ($r < -0.83$), which is expected given its formulation as an error-based metric. These pixel-level metrics also exhibit moderate correlations with segmentation-based measures such as Dice and IoU (typically $0.5 < r < 0.7$). This suggests that improvements in image fidelity may be associated with corresponding gains in staining accuracy; however, the moderate strength of these correlations indicates that this relationship should be interpreted cautiously and not inferred as strongly predictive.

Segmentation metrics exhibit strong internal consistency. Dice and IoU scores across the DAPI, CD3, and PanCK channels are highly correlated ($r > 0.9$), indicating that they capture similar aspects of overlap-based performance. In contrast, Hausdorff Distance (HD) shows strong negative correlations with Dice and IoU ($r < -0.8$), reflecting its sensitivity to boundary-level localization errors.

A particularly notable observation is the behavior of FID. Compared with both pixel-level and segmentation-based metrics, FID exhibits relatively weak and inconsistent correlations (typically $|r| < 0.4$). This suggests that improved distribution alignment does not necessarily correspond to improved staining accuracy or spatial localization. In practice, methods that achieve favorable FID scores may still exhibit substantial errors in recovering the spatial distribution of target staining patterns.

Overall, the correlation analysis reveals three largely distinct groups of metrics. Pixel-level metrics primarily measure reconstruction fidelity, segmentation metrics quantify staining accuracy and spatial correspondence, and distribution-based metrics assess global appearance similarity. These findings reinforce the importance of evaluating virtual staining models using complementary metrics that capture both image quality and staining accuracy, rather than relying exclusively on image-fidelity or distribution-based measures \cite{liu2022bci,li2023adaptive,dubey2024vims,hu2025mcs,hussain2026vit,bian2024hemit}. This correlation analysis provides further evidence that the multi-faceted evaluation strategy used in our manuscript offers a more comprehensive and complete characterization of the virtual staining model’s performance at test time.

\subsection{Ablation Experiments}
We conduct a comprehensive ablation study to evaluate the key design choices underlying \name, including the feature extraction backbone (convolutional, Swin Transformer, or hybrid), neighborhood window size, channel capacity, and the contribution of perceptual loss (including its weighting and the number of feature encoders employed). These experiments quantify the impact of each component on overall model performance. To reduce computational cost and ensure a controlled comparison, all ablations—except the backbone analysis—are performed using the convolution-only variant of \name, providing a consistent setting for isolating the effect of individual design choices.

We further evaluate the robustness of \name under limited-data conditions by training on progressively smaller subsets of patients. These experiments assess the extent to which the proposed framework can maintain performance when the amount of available training data is reduced. In addition, we investigate resolution scalability by training the model at lower image resolutions and performing inference at higher resolutions. This experiment evaluates whether \name can learn useful representations from lower-resolution inputs while preserving performance during high-resolution inference. The primary motivation for this analysis is not data scarcity, but rather the potential computational benefits of low-resolution training, which can substantially reduce memory and training costs while retaining competitive performance.

\textbf{Backbone Architecture Ablation.} Tables~\ref{tab:quantiative_metrics_ablation_hemit_backbone} and supplementary table \ref{tab:quantiative_metrics_ablation_CD3_backbone} present an ablation study evaluating the impact of different encoder backbones in \name. We compare three configurations: a convolution-only backbone, a Swin Transformer-only backbone, and a hybrid combination of both.
\begin{table*}[!thb]
\setlength{\tabcolsep}{2.4pt}
\scalebox{0.85}{
    \begin{tabular}{c||ccc||ccc||ccc|ccc|ccc|c}
   & \multicolumn{3}{c||}{\bf Texture Metrics } &  \multicolumn{3}{c||}{\bf Distribution Metrics} & \multicolumn{9}{c|}{\bf Segmentation Metrics}& infer\\ 
   & \multicolumn{3}{c||}{ } &  \multicolumn{3}{c||}{} & \multicolumn{3}{c|}{\it DAPI} & \multicolumn{3}{c|}{\it CD3} &  \multicolumn{3}{c|}{\it PanCK}& time\\  
   Methods & PSNR $\uparrow$ & SSIM $\uparrow$ & MSE$\downarrow$& FID $\downarrow$ & Prec $\uparrow$ & Recall $\uparrow$&  Dice $\uparrow$ & IoU $\uparrow$ & HD $\downarrow$ &  Dice $\uparrow$ & IoU $\uparrow$ & HD $\downarrow$ &  Dice $\uparrow$ & IoU $\uparrow$ & HD $\downarrow$ & (sec)\\
    \hline
    \hline
          {\name }   &  \bf 32.26 & \bf 0.8450 & \bf 55.38 & \bf 12.01 & \bf 0.8685 & \bf 0.7776 & \bf 0.6355 &  0.5496 & \bf 43.78 &  0.5026 &  0.4125 & 58.30 & \bf 0.8538 & \bf 0.7713 & \bf 12.58 & 0.129  \\
          
             {\name} \{Conv\}   & 31.06 & 0.7959 & 65.12 &  16.01 & 0.8154 &  0.7240 &  0.6191 &  0.5251 & 51.27 & \bf  0.5336 & \bf  0.4409 & \bf 57.40 & 0.8458 & 0.7585 & 13.93 & \bf 0.044 \\
               {\name }  \{Swin\} & 29.32	& 0.8069	& 69.76	& 27.77	& 0.8302 & 0.5618	& 0.5853	& 0.4884	& 53.18	& 0.3364	& 0.2468	& 76.58 & 0.8243 & 0.7304 &12.96 & 0.073 \\
    \end{tabular}}
  \caption{\textbf{Backbone Ablation on the HEMIT Dataset.} Quantitative comparison of a convolution-only backbone, a Swin Transformer-only backbone, and the proposed hybrid convolution--Swin architecture. Inference time is reported in seconds per $256 \times 256$ image patch using a batch size of one. The hybrid architecture achieves the strongest overall performance across most image-fidelity, distribution-based, and segmentation metrics, particularly for the DAPI and PanCK channels. An exception is the CD3 channel, where the convolution-only backbone attains slightly higher Dice and IoU scores. These results suggest that different staining targets may benefit from different balances of local and global contextual information, while also highlighting the complementary strengths of convolutional and transformer-based representations.
}
    \label{tab:quantiative_metrics_ablation_hemit_backbone}
\end{table*}

Across all settings, \name maintains competitive performance across the backbone variants reported in Tables~\ref{tab:quantiative_metrics_ablation_hemit_backbone} and \ref{tab:quantiative_metrics_ablation_CD3_backbone}, indicating that the overall framework is relatively robust to the choice of feature extraction architecture. Nevertheless, the hybrid convolution--Swin backbone achieves the strongest overall performance across the majority of image-fidelity and segmentation-based metrics. On the HEMIT dataset, the hybrid model attains the best performance for the DAPI and PanCK channels, as well as the strongest overall texture and distribution-based metrics. An interesting exception is the CD3 channel, where the convolution-only backbone achieves slightly higher Dice and IoU scores than the hybrid model. In contrast, on the CD3 IHC dataset, the hybrid architecture consistently outperforms both the convolution-only and Swin-only variants across most metrics. These observations suggest that the relative contribution of local and global contextual information may vary across staining tasks and imaging modalities. While additional studies are needed to better understand these differences, the results indicate that combining convolutional and transformer features provides a more effective representation for most virtual staining tasks.

Importantly, the improved performance of the hybrid backbone comes with increased computational cost. As shown in Table~\ref{tab:quantiative_metrics_ablation_hemit_backbone}, the convolution-only and Swin-only variants achieve inference times of 0.044 and 0.073 seconds per $256\times256$ image patch, respectively, compared with 0.129 seconds for the hybrid model (batch size = 1). Although the hybrid architecture is slower, all variants remain computationally practical, with inference times corresponding to the order of minutes per whole-slide image depending on slide size and available hardware resources. Consequently, the improved quantitative performance achieved by the hybrid backbone represents a reasonable trade-off for the additional computational cost. Further optimization of the architecture to improve inference efficiency while preserving performance remains an important direction for future work.

Overall, these results suggest that convolutional and transformer features provide complementary information. Convolutional layers capture fine-grained local morphology, whereas transformer layers model broader contextual relationships and long-range dependencies. Integrating both representations yields richer pixel-level features and consistently improves performance across multiple virtual staining tasks, supporting the use of hybrid architectures for accurate and robust virtual staining.

\textbf{Ablation without Perceptual Loss.} To evaluate the contribution of perceptual supervision, we train variants of \name with all perceptual loss terms removed. The corresponding results are reported in supplementary table~\ref{tab:without_perceptual_loss}. Across both the CK8/18 and CD3 datasets, removing the perceptual loss results in performance degradation across all metric groups. The largest decline is observed in the distribution-based metrics, where FID increases substantially (e.g., from 17.69 to 116.16 on CK8/18 and from 33.79 to 101.61 on CD3), accompanied by pronounced reductions in precision and recall. In contrast, texture metrics (PSNR, SSIM, and MSE) and segmentation-based metrics exhibit smaller, though still measurable, performance drops.

These results suggest that perceptual supervision plays an important role in aligning generated images with the feature distribution of real stains, thereby improving visual quality and distributional consistency. At the same time, the relatively smaller changes observed in segmentation-based metrics indicate that staining accuracy is influenced not only by perceptual supervision but also by the underlying architecture and pixel-level learning objective. Overall, the results demonstrate that perceptual loss contributes meaningfully to both image realism and staining performance, with its most pronounced impact occurring in distribution-based evaluation metrics.

Qualitative results in Fig.~\ref{fig:qualitative_without_perceptual_loss} further support these quantitative observations. For both the CK8/18 and CD3 datasets, removing the perceptual loss produces visibly smoother outputs with reduced contrast and less distinct structural detail compared with the full model. Fine-grained characteristics such as cell boundaries, local staining intensity variations, and tissue texture patterns are less clearly preserved, resulting in images that deviate more noticeably from the appearance of the reference IHC stains. These observations suggest that pixel-level supervision alone is insufficient to fully capture the visual characteristics of real stained images.

Interestingly, despite the reduction in visual quality, many stain-positive regions remain approximately localized. This observation is consistent with the quantitative results, where segmentation-based metrics decline less dramatically than distribution-based metrics. Together, these findings suggest that coarse spatial localization can largely be learned from the pixel-level objective, whereas accurate reconstruction of stain appearance, local texture, and intensity variation benefits substantially from perceptual supervision. Overall, the results indicate that perceptual loss plays an important role in improving both visual fidelity and distributional alignment, enabling the generation of sharper and more realistic virtual stains.

\begin{figure}[!htb]
    \centering
    \includegraphics[trim={0.0cm 6.7cm 18.0cm 0.0cm}, clip=true, width=1.0\linewidth]{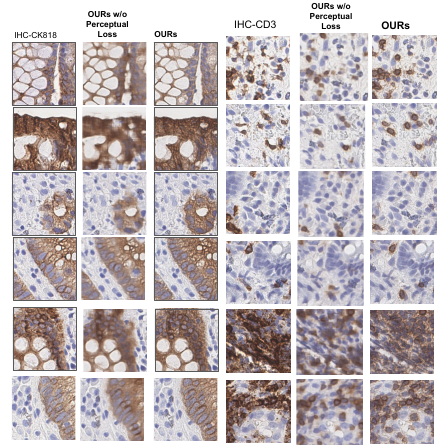}
   \caption{\textbf{Effect of Perceptual Loss on Virtual IHC Generation.} Qualitative comparison of virtual IHC images generated with and without perceptual loss. Perceptual supervision improves visual fidelity by enhancing local contrast, preserving fine structural details, and producing staining patterns that more closely resemble the corresponding reference IHC images.
}
     \label{fig:qualitative_without_perceptual_loss}
\end{figure}

\textbf{$\lambda$ Value Ablation.} To evaluate the impact of perceptual supervision, we vary both the number of perceptual encoders and the corresponding loss weights $(\lambda_1,\lambda_2,\lambda_3)$, as summarized in supplementary Table~\ref{tab:quantiative_metrics_lambda}. Accordingly, the total loss function was modified as follows.
\begin{align*}
    \mathcal{L}_{total}=
\mathcal{L}_{Imp}+\lambda_1 \mathcal{L}_{perceptual}(\text{AlexNET}) \\ + \lambda_2 \mathcal{L}_{perceptual}(\text{VGG}) \\+ \lambda_3 \mathcal{L}_{perceptual}(\text{ResNET50}) 
\end{align*}

Several observations emerge from this study:

\begin{itemize}
\item Incorporating perceptual supervision substantially improves distribution-based metrics relative to training without perceptual loss. In particular, FID, precision, and recall improve markedly once perceptual features are introduced.

\item Texture-based metrics (PSNR, SSIM, and MSE) achieve their strongest values when perceptual loss is removed. This behavior is expected, as the model focuses exclusively on optimizing pixel-level reconstruction in the absence of feature-level constraints.

\item Increasing the number of perceptual encoders or scaling the corresponding loss weights does not consistently improve performance. For example, moving from $(1,1,0)$ to $(10,10,0)$ or $(100,100,0)$ yields little benefit and occasionally degrades performance.

\item The effect of perceptual supervision varies across staining channels. DAPI generally benefits from perceptual loss, whereas CD3 exhibits its strongest segmentation performance under weaker perceptual supervision. PanCK performance is comparatively stable across a range of configurations.

\item Among the evaluated settings, $(\lambda_1,\lambda_2,\lambda_3)=(1,1,0)$ provides the most balanced trade-off between image fidelity, distributional alignment, and segmentation accuracy. The configuration $(0.1,0.1,0)$ achieves comparable performance and yields the best overall FID score.

\item An interesting observation is the apparent trade-off between channels: configurations that improve CD3 performance occasionally reduce DAPI performance, suggesting that optimization for different markers may compete for representational capacity within the shared model.

\end{itemize}

\textbf{Channel Dimension $C$ Ablation.} To determine the optimal representation dimension C, we conducted an ablation study varying the channel size, as reported in supplementary supplementary Table~\ref{tab:quantiative_channel_dimension}. Performance improves with increasing capacity up to C=64, beyond which gains plateau or degrade. Accordingly, we use C=64 for all reported results.

\textbf{Window Size Ablation ($w$).} To evaluate the impact of the neighborhood window size, we conduct an ablation study using $w \in {3,5,7}$, with results summarized in Table~\ref{tab:quantiative_window_size_dimension}. Increasing the window size generally improves performance across image-fidelity, distribution-based, and segmentation metrics, with $w=7$ achieving the strongest overall results. However, the magnitude of these improvements is relatively modest. For example, the Dice score for the DAPI channel increases from 0.6034 ($w=3$) to 0.6053 ($w=7$), while the PanCK Dice score increases from 0.8310 to 0.8353.

In contrast, the computational cost increases substantially. Training with $w=3$ requires 6.91 GB of GPU memory, whereas $w=7$ requires 35.54 GB—more than a fivefold increase. This significantly affects training efficiency and scalability, particularly for high-resolution pathology images. For the hybrid convolution--Swin architecture, the increase in computational cost is even more pronounced. Consequently, although larger windows provide measurable improvements, the gains in staining accuracy are relatively small compared with the associated increase in memory and computational requirements. Based on this trade-off, we adopt $w=3$ as the default setting throughout the paper. This configuration provides competitive performance while substantially reducing computational cost. Nevertheless, larger window sizes may be beneficial when computational resources are less constrained or when applications require broader contextual information.

\textbf{Coordinate Encoding Ablation.} To isolate the effect of positional encoding from model capacity, we perform this ablation using a smaller convolutional backbone (1.7M parameters) rather than the full convolutional architecture (39M parameters). Table~\ref{tab:quantiative_metrics_coordinate_encoding_ablation} compares three coordinate representations: raw spatial coordinates, Fourier embeddings commonly used in implicit neural representations~\cite{mildenhall2021nerf}, and the proposed learned coordinate encoding.

Overall, raw coordinates and learned encodings achieve similar performance across most metrics, whereas Fourier encoding consistently underperforms. In particular, Fourier encoding yields lower image-fidelity scores, worse distribution-based metrics, and reduced segmentation accuracy, suggesting that high-frequency positional embeddings may be less suitable for this virtual staining setting.

Comparing raw and learned coordinate representations, the learned encoding achieves slightly stronger image-fidelity and distribution-based performance, obtaining the highest PSNR, lowest MSE, best FID, and highest recall. For segmentation-based evaluation, the learned encoding performs best on the DAPI channel across Dice, IoU, and Hausdorff Distance, while the raw coordinate representation achieves slightly better performance on the CD3 channel. For the PanCK channel, both approaches produce nearly identical results, differing only marginally across metrics.

Taken together, these results indicate that the framework is relatively robust to the choice of coordinate representation. Nevertheless, the learned coordinate encoding provides the most balanced overall performance across image-fidelity, distribution-based, and segmentation metrics. Consequently, we adopt the learned coordinate encoding as the default configuration in all subsequent experiments.

\begin{table*}[!thb]
\setlength{\tabcolsep}{2.5pt}
\scalebox{0.85}{
    \begin{tabular}{p{0.05cm}|c||ccc||ccc||ccc|ccc|ccc}
   & & \multicolumn{3}{c||}{\bf Texture Metrics } &  \multicolumn{3}{c||}{\bf Distribution Metrics} & \multicolumn{9}{c}{\bf Segmentation Metrics}\\ 
   & & \multicolumn{3}{c||}{ } &  \multicolumn{3}{c||}{} & \multicolumn{3}{c|}{\it DAPI} & \multicolumn{3}{c|}{\it CD3} &  \multicolumn{3}{c}{\it PanCK}\\  
   & Methods & PSNR $\uparrow$ & SSIM $\uparrow$ & MSE$\downarrow$& FID $\downarrow$ & Prec $\uparrow$ & Recall $\uparrow$&  Dice $\uparrow$ & IoU $\uparrow$ & HD $\downarrow$ &  Dice $\uparrow$ & IoU $\uparrow$ & HD $\downarrow$ &  Dice $\uparrow$ & IoU $\uparrow$ & HD $\downarrow$\\
    \hline
    \hline
          
           &  Raw Coordinates   & 30.85 & \bf 0.7938 & 66.16 & 17.00 &  0.8200 & 0.7055 & 0.6138 & 0.5193 & 51.89 & \bf 0.5061 & \bf 0.4126 & \bf 62.07 & \bf 0.8458 & \bf 0.7591 & \bf 14.22 \\
             & Fourier Encoding & 30.71 & 0.7663 & 71.15 & 30.95 & 0.7203 & 0.6745 & 0.6130 & 0.5202 & 55.34 & 0.4759 & 0.3889 & 68.62 & 0.8279 & 0.7405 & 19.70 \\
             &   Learned  Encoding(\textbf{Ours}) &  \bf 30.86 & 0.7910 & \bf 65.93 &  \bf 16.49 & \bf 0.8224 &  \bf 0.7163 &  \bf 0.6147 &  \bf 0.5203 & \bf 51.56 &  0.5031 &  0.4107 & 62.57 & \bf 0.8458 &  0.7587 &  14.25 \\
    \end{tabular}}
 \caption{\textbf{Coordinate Encoding Ablation on HEMIT.} Quantitative comparison of raw spatial coordinates, Fourier embeddings~\cite{mildenhall2021nerf}, and the learned coordinate encoding used in \name. Raw coordinates and learned encodings achieve broadly similar performance across most evaluation metrics, whereas Fourier encoding generally yields lower performance. The learned encoding provides a slightly more favorable overall balance across image-fidelity, distribution-based, and segmentation metrics and is therefore used as the default coordinate representation in subsequent experiments.
}
    \label{tab:quantiative_metrics_coordinate_encoding_ablation}
\end{table*}

 \textbf{Ablations with Limited Training Data.} The results in Table~\ref{tab:quantiative_metrics_ablations_dataset_size} indicate that \name degrades gracefully as the amount of training data is reduced. Even when trained on only $25\%$ or $10\%$ of the original training set, \name remains highly competitive and consistently achieves the strongest segmentation-based performance across most evaluation metrics. In particular, overlap-based measures such as Dice and IoU remain substantially higher than those of competing methods, indicating that the model retains strong staining localization capabilities even under limited-data conditions.

Comparisons with baseline methods trained on the same reduced datasets reveal two notable trends. First, \name continues to outperform competing approaches on most segmentation metrics, despite a reduction in available training data. Second, unpaired methods exhibit a more pronounced performance decline than paired approaches, suggesting that direct supervision becomes increasingly important as the amount of training data decreases. While paired methods such as PyramidPix2Pix remain competitive on several image-fidelity metrics, \name consistently achieves stronger staining-accuracy performance across both low-data settings.

One possible explanation for this behavior is the pixel-level formulation of \name, which provides dense supervision throughout training and enables the model to learn from a large number of spatially localized examples. This may improve sample efficiency relative to methods that rely more heavily on patch-level image generation objectives. Nevertheless, performance still decreases as the training set becomes smaller, suggesting that the feature extraction backbone remains a limiting factor in low-data regimes. Future work could therefore explore transfer learning, domain adaptation, or pretrained feature extractors to further improve data efficiency and robustness.

Overall, these results suggest that \name maintains strong performance under substantial reductions in training data and remains particularly effective at preserving staining accuracy in low-data settings.

 \begin{table*}[!thb]
\setlength{\tabcolsep}{2.5pt}
\scalebox{0.85}{
    \begin{tabular}{p{1.2cm}|c||ccc||ccc||ccc|ccc|ccc}
  Data Ratio  & & \multicolumn{3}{c||}{\bf Texture Metrics } &  \multicolumn{3}{c||}{\bf Distribution Metrics} & \multicolumn{9}{c}{\bf Segmentation Metrics}\\ 
  & & \multicolumn{3}{c||}{ } &  \multicolumn{3}{c||}{} & \multicolumn{3}{c|}{\it DAPI} & \multicolumn{3}{c|}{\it CD3} &  \multicolumn{3}{c}{\it PanCK}\\  
   & Methods & PSNR $\uparrow$ & SSIM $\uparrow$ & MSE$\downarrow$& FID $\downarrow$ & Prec $\uparrow$ & Recall $\uparrow$&  Dice $\uparrow$ & IoU $\uparrow$ & HD $\downarrow$ &  Dice $\uparrow$ & IoU $\uparrow$ & HD $\downarrow$ &  Dice $\uparrow$ & IoU $\uparrow$ & HD $\downarrow$\\
    \hline
    \hline

      1.0   &  {\name }   &   32.26 &  0.8450 &  55.38 &  12.01 &  0.8685 & {0.7776} &  0.6355 & 0.5496 &  43.78 &  0.5026 &  0.4125 &  58.30 &  0.8538 &  0.7713 &  12.58 \\
      \hline
       0.25    & {\name}   &  \bf 31.36 & 0.8405 & 59.12 & 14.99 & \bf 0.8587 & 0.7177 & \bf 0.6229 & \bf 0.5319 & \bf 45.98 & \bf 0.4581 & \bf 0.3674 & \bf 71.99 & \bf 0.8481 & \bf 0.7631 & \bf 11.97 \\
       0.25 & Pix2Pix & 28.75 & 0.8410 & 62.63 & 20.97 & 0.7411 & 0.5083 & 0.5610 & 0.4726 & 58.19 & 0.2603 & 0.1802 & 87.27 & 0.8148 & 0.7213 & 12.78 \\
       0.25 & PyrmaidP2P & 30.61 & 0.8725 & \bf 53.58 & \bf 10.07 & 0.8488 & \bf 0.8114 & 0.5973 & 0.5105 & 42.69 & 0.3881 & 0.2974 & 63.19 & 0.8317 & 0.7458 & 12.10 \\
       0.25 & CycleGAN & 21.82 & 0.5576 & 148.38 & 32.41 & 0.7393 & 0.3246 & 0.2844 & 0.2054 & 98.23 & 0.0294 & 0.0144 & 100.99 & 0.0269 & 0.0159 & 67.91 \\
       0.25 & QS-GAN & 25.14 & 0.7072 & 90.91 & 120.99 & 0.6910 & 0.0071 & 0.0752 & 0.0463 & 140.99 & 0.0234 & 0.0134 & 134.79 & 0.4816 & 0.3265 & 55.69 \\
       \hline
       0.1   & {\name }   & \bf 30.22 &  0.8231 &  63.40 & 21.58 & {0.8009} & 0.6486 & \bf 0.5959 & \bf 0.5003 & 49.26 & \bf 0.3892 & \bf 0.2969 & 79.50 & \bf 0.8354 & \bf 0.7463 & \bf 12.13 \\
       0.1 & Pix2Pix & 27.42 & 0.4576 & 137.35 & 96.76 & 0.4642 & 0.0199 & 0.6040 & 0.5252 & 48.94 & 0.0674 & 0.0381 & 111.68 & 0.8084 & 0.7008 & 14.08 \\
       0.1 & PyramidP2P & 29.79 & \bf 0.8605 & \bf 58.11 & 10.72 & 0.8196 & 0.7674 & 0.5745 & 0.4844 & \bf 48.22 & 0.3601 & 0.2687 & \bf 67.95 & 0.8288 & 0.7393 & 12.27 \\
       0.1 & CycleGAN & 22.27 & 0.5753 & 145.96 & 19.52 & 0.7491 & 0.4293 & 0.3046 & 0.2188 & 94.22 & 0.0201 & 0.0107 & 110.63 & 0.0263 & 0.0140 & 66.79 \\
       0.1 & QS-GAN & 20.11 & 0.5658 & 137.70 & 101.55 & 0.6049 & 0.0062 & 0.0574 & 0.0322 & 158.84 & 0.1592 & 0.0955 & 105.82 & 0.2981 & 0.1784 & 43.84 \\
    \end{tabular}}
  \caption{\textbf{Training Set Size Ablation on the HEMIT Dataset.} Performance of \name and representative paired and unpaired baselines when trained using reduced fractions of the original training set, while keeping the test set unchanged. As training data become more limited, competing methods remain competitive on some image-fidelity metrics, particularly PyramidPix2Pix. However, \name consistently achieves the strongest segmentation-based performance, indicating improved preservation of staining patterns and spatial localization under low-data training conditions.}
    \label{tab:quantiative_metrics_ablations_dataset_size}
\end{table*}



\textbf{Resolution-Flexible Training and Inference.} Because \name is formulated as a coordinate-conditioned implicit model, it naturally supports training and inference at different spatial resolutions. Coordinate inputs are fundamental to the proposed formulation; removing them would eliminate the implicit representation mechanism and therefore constitute a substantially different model. For this reason, we do not include an ablation without coordinate conditioning.

To evaluate the resolution-flexible properties of the framework, we train \name on low-resolution image tiles ($64\times64$) obtained by downsampling the original $256\times256$ training patches. During inference, predictions are generated at the original resolution by querying the model using high-resolution spatial coordinates. This setting substantially reduces computational requirements. For example, GPU memory consumption decreases from approximately 132 GB to 12 GB when training the Swin-based model with the same batch size.

The quantitative results in Tables~\ref{tab:quantiative_metrics_ablations_lr} and \ref{tab:quantiative_metrics_ablations_lr_CD3} show a consistent reduction in performance across most metrics when training is performed at low resolution. The largest degradation is observed in distribution-based metrics and several segmentation measures, indicating that low-resolution training limits the model's ability to capture fine-grained spatial information present in the original data. Nevertheless, the performance reduction is moderate for many staining-accuracy metrics, suggesting that the underlying stain translation mapping is still learned reasonably well.

Qualitative results (Fig.~\ref{fig:low_resolution_inference}) provide additional insight. Despite being trained exclusively on low-resolution images, the model often identifies and highlights many of the same stain-positive regions as the full-resolution model. However, the resulting images exhibit reduced detail and lower visual fidelity, particularly in regions containing fine cellular structures. These observations suggest that while the semantic mapping between source and target domains is largely preserved, recovering high-frequency details remains challenging when training is performed at substantially lower resolutions.

Overall, these experiments demonstrate that \name retains the ability to generate plausible high-resolution outputs even when trained at lower resolutions. This capability provides a practical trade-off between computational efficiency and predictive performance, which may be particularly useful in large-scale digital pathology workflows. Future work could investigate resolution-aware training strategies, such as low-resolution pretraining followed by high-resolution fine-tuning, as well as alternative positional encoding schemes designed to improve generalization across resolutions.

 \begin{table*}[!thb]
\setlength{\tabcolsep}{2.5pt}
\scalebox{0.85}{
    \begin{tabular}{p{0.1cm}|c||ccc||ccc||ccc|ccc|ccc}
   & & \multicolumn{3}{c||}{\bf Texture Metrics } &  \multicolumn{3}{c||}{\bf Distribution Metrics} & \multicolumn{9}{c}{\bf Segmentation Metrics}\\ 
   & & \multicolumn{3}{c||}{ } &  \multicolumn{3}{c||}{} & \multicolumn{3}{c|}{\it DAPI} & \multicolumn{3}{c|}{\it CD3} &  \multicolumn{3}{c}{\it PanCK}\\  
   & Methods & PSNR $\uparrow$ & SSIM $\uparrow$ & MSE$\downarrow$& FID $\downarrow$ & Prec $\uparrow$ & Recall $\uparrow$&  Dice $\uparrow$ & IoU $\uparrow$ & HD $\downarrow$ &  Dice $\uparrow$ & IoU $\uparrow$ & HD $\downarrow$ &  Dice $\uparrow$ & IoU $\uparrow$ & HD $\downarrow$\\
    \hline
    \hline
       &  {\name }   &  32.26 &  0.8450 & 55.38 &  12.01 &  0.8685 & 0.7776 & 0.6355 &  0.5496 &  43.78 &  0.5026 &  0.4125 &  58.30 &  0.8538 &  0.7713 & 12.58 \\
&  {{\sc ImplicitStainer-LR}}   & 30.75 & 0.8465 & 59.91 & 73.83 & 0.2140 & 0.2140 & 0.5877 & 0.5021 & 44.67 & 0.3988 & 0.3012 & 68.85 & 0.8128 & 0.7096 & 12.40 \\
    \end{tabular}}
  \caption{\textbf{Low-Resolution Training and High-Resolution Inference on HEMIT.} Quantitative comparison between \name trained directly on $256\times256$ images and {\sc ImplicitStainer-LR}, which is trained on downsampled $64\times64$ images but evaluated at $256\times256$ resolution through coordinate-based inference. Despite the substantial reduction in training resolution, {\sc ImplicitStainer-LR} retains competitive performance on several staining-accuracy metrics, demonstrating the resolution-flexible nature of the proposed implicit formulation.
}
    \label{tab:quantiative_metrics_ablations_lr}
\end{table*}

\textbf{Failure Modes Across Different Methods.} During qualitative analysis, we identified several challenging examples for which most methods exhibit degraded performance. Representative failure cases are shown in Fig.~\ref{fig:qualitative_results_add_failure}. For the CD3 dataset, the first two rows exhibit widespread understaining across many models, whereas the remaining examples show a tendency toward overstaining. One possible explanation is the visual similarity between CD3-positive lymphocytes and other morphologically similar lymphoid cells that do not express CD3, making accurate localization difficult from H\&E appearance alone. These examples highlight the challenges associated with inferring marker-specific expression from subtle morphological cues.

For the CK8/18 dataset, failure cases are more frequently associated with noisy image regions or reduced tissue quality. Relative to CD3, CK8/18 appears to be a comparatively easier staining task, and most methods produce visually plausible outputs in the majority of cases. Nevertheless, false-positive staining artifacts are occasionally observed, particularly for Pix2Pix, suggesting sensitivity to image quality and local staining variability.

For the HEMIT dataset, many failure cases correspond to images with atypical color characteristics, including a pronounced pink hue that differs from the appearance of most H\&E samples in the dataset. These examples also often exhibit weaker visual correspondence between the H\&E image and the target mIF channels. Such cases may arise from factors including staining variability, image noise, registration inaccuracies, or other dataset-specific inconsistencies. Regardless of the underlying cause, these samples appear substantially more difficult for all evaluated methods, resulting in reduced performance across the board.

These observations suggest several potential directions for future improvement, including stain normalization, stronger color augmentation strategies, and improved data curation procedures to identify noisy or poorly aligned samples. However, a systematic investigation of these factors is beyond the scope of the present work and remains an important direction for future research.

\subsection{Limitations and Future Work Directions}

Although \name demonstrates strong performance on paired datasets, several limitations remain. First, the current formulation assumes spatial correspondence between source and target domains and therefore relies on paired, registered training data. As a result, when applied to widely used datasets containing adjacent but misaligned tissue sections (e.g., BCI~\cite{liu2022bci} and MIST~\cite{li2023adaptive}), preliminary experiments indicate reduced performance relative to methods specifically designed for unpaired image translation. Extending the framework to handle spatial misalignment therefore represents an important direction for future work. Potential approaches include jointly learning registration and translation, incorporating learnable spatial offsets (e.g., $(\Delta x,\Delta y)$), modeling deformation fields~\cite{xia2022vision}, or leveraging domain adaptation techniques based on pretrained image translation models~\cite{alijani2024vision,kataria2024unsupervised}. Such extensions could substantially broaden the applicability of the framework to weakly paired or unpaired virtual staining settings.

Second, \name incurs higher training costs than many existing virtual staining methods due to its hybrid convolution--transformer backbone and pixel-level prediction formulation. Training the full model for 200 epochs on a single GPU requires approximately eight days, compared with roughly three to six days for paired methods such as PyramidPix2Pix and AdaptiveNCE, although inference times remain comparable. As demonstrated in our resolution ablation experiments, lower-resolution training can substantially reduce computational requirements but currently results in reduced performance. Improving training efficiency while preserving spatial accuracy therefore remains an important challenge. Future work will investigate more efficient attention mechanisms, including axial attention~\cite{ho2019axial,bin2025efficientmorph}, deformable attention~\cite{xia2022vision}, and FlashAttention~\cite{dao2022flashattention}, which may enable more efficient modeling of long-range dependencies while reducing memory consumption and training time.

Finally, while the proposed framework supports resolution-flexible inference, our experiments show that training at substantially lower resolutions still leads to measurable performance degradation. Developing resolution-aware training strategies, such as low-resolution pretraining followed by high-resolution fine-tuning, may help further improve scalability while retaining the advantages of the implicit formulation.

\section{Conclusion}

In this work, we introduced \name, a coordinate-conditioned, pixel-level framework for virtual staining based on neural implicit representations. By formulating virtual staining as a deterministic regression problem, \name supports resolution-flexible training and inference while producing consistent outputs for a given input image. Comprehensive evaluations across multiple virtual staining datasets demonstrate that \name achieves competitive image-fidelity performance and strong staining accuracy, while remaining effective under reduced training-data regimes. In addition, our analysis highlights the importance of combining image-fidelity, distribution-based, and staining-accuracy metrics when evaluating virtual staining models, as different metric families capture complementary aspects of performance. We further showed that training at lower resolutions and performing inference at higher resolutions remains feasible within the proposed framework, providing a practical trade-off between computational efficiency and predictive performance. Although this setting leads to some loss of fine-grained detail, the model continues to recover many of the key staining patterns present in the target images. At the same time, our experiments identify several limitations, including reduced performance on unpaired datasets, increased training costs associated with the hybrid convolution--transformer backbone, and challenges arising from ambiguous cellular structures and noisy inputs.

Future work will focus on extending the framework to weakly paired and unpaired settings, improving training efficiency through more computationally efficient architectures, and exploring applications beyond virtual staining, including other medical image translation tasks. We also plan to investigate transfer learning, few-shot adaptation, and resolution-aware training strategies to further improve data efficiency and scalability.

\section*{CRediT Authorship Contribution Statement}
\textbf{Tushar Kataria}; Writing – original draft, Writing – review \& editing, Visualization, Validation, Software, Methodology, Investigation, Formal analysis, Data curation, Conceptualization. \textbf{Beatrice S. Knudsen}: Writing
– review \& editing, Supervision, Resources, Project administration,
Funding acquisition, \textbf{Shireen Y Elhabian}: Writing – review \& editing, Supervision, Resources, Project administration, Funding acquisition, Conceptualization.

\section*{Declaration of competing interest}
The authors declare that they have no known competing financial interests or personal relationships that could have appeared to influence the work reported in this paper.

\section*{Acknowledgments}
The National Science Foundation supported this work under grant number NSF2217154. We also thank the Department of Pathology, the Kahlert School of Computing and Scientific Computing and Imaging Institute at the University of Utah for their support of this project. We also gratefully acknowledge the Center for High Performance Computing at the University of Utah for providing the computational resources used in this work. These resources were partially supported by the NIH Shared Instrumentation Grant 1S10OD021644-01A1. The content is solely the authors’ responsibility and does not necessarily represent the official views of the National Science Foundation. 

\section*{Declaration of generative AI and AI-assisted technologies in the manuscript preparation process.}
During the preparation of this manuscript, the author(s) used ChatGPT to check grammar and improve sentence clarity and readability. After using this tool, the author(s) carefully reviewed and edited the content as needed and take full responsibility for the content of the published article. All the figures were manually create using power point or google slides or python libraries, with only the 3D representation of the feature space($\Psi^{swin},\Psi^{conv}$) being created using ChatGPT in Figure \ref{fig:pipeline_fig}. 




\bibliographystyle{model2-names.bst}\biboptions{authoryear}
\bibliography{refs}
\newpage
\onecolumn
\centering
\section*{Supplementary Material}
\setcounter{figure}{0}
\renewcommand{\thefigure}{SF\arabic{figure}}
\setcounter{section}{0}
\renewcommand{\thesection}{S\arabic{section}}
\setcounter{table}{0}
\renewcommand{\thetable}{ST\arabic{table}}
\setcounter{page}{1}

\begin{table*}[!htb]
\centering
\begin{tabular}{p{1.5cm}| p{1.1cm} | p{2.0cm} | p{12.0cm}}
\hline
\textbf{Category} & \textbf{Metric} & \textbf{Direction} & \textbf{Description} \\
\hline
Image & PSNR & $\uparrow$ & Measures reconstruction quality based on pixel-wise intensity differences. \\
Realism & SSIM & $\uparrow$ & Evaluates structural similarity between images considering luminance, contrast, and structure. \\
& MSE & $\downarrow$ & Computes average squared difference between pixel intensities. \\
& FID & $\downarrow$ & Measures distance between real and generated image distributions in feature space. \\
& KID & $\downarrow$ & Kernel-based distance between feature distributions, similar to FID but unbiased. \\
& Precision & $\uparrow$ & Fraction of generated samples within the real data distribution. \\
& Recall & $\uparrow$ & Fraction of real samples covered by the generated distribution. \\
\hline
& Dice & $\uparrow$ & Measures overlap between predicted and ground-truth masks. \\
Staining  & IoU & $\uparrow$ & Intersection over Union; ratio of overlap to union between masks. \\
Accuracy & HD & $\downarrow$ & Hausdorff Distance; measures boundary mismatch. \\
& TPR & $\uparrow$ & True Positive Rate; correctly identified positive pixels. \\
& TNR & $\uparrow$ & True Negative Rate; correctly identified negative pixels. \\
\hline
\end{tabular}
\caption{\textbf{Evaluation Metrics Used.} Metrics for image realism and staining accuracy.}
\label{tab:metric_table}
\end{table*}

\begin{table*}[!htb]
\centering
\scriptsize
\setlength{\tabcolsep}{3pt}
\renewcommand{\arraystretch}{1.1}
\resizebox{\textwidth}{!}{%
\begin{tabular}{lcccccccc}
\toprule
Method & PSNR & SSIM & MSE & Dice & Jaccard & HD & TPR & TNR \\
\midrule

CycleGAN & \mstd{19.90}{3.89} & \mstd{0.58}{0.12} & \mstd{232.79}{38.82} & \mstd{0.62}{0.32} & \mstd{0.52}{0.30} & \mstd{41.28}{49.75} & \mstd{0.68}{0.34} & \mstd{0.91}{0.11} \\
UNIT & \mstd{12.36}{1.77} & \mstd{-0.00}{0.17} & \mstd{296.93}{30.11} & \mstd{0.54}{0.37} & \mstd{0.46}{0.34} & \mstd{50.27}{42.36} & \mstd{0.57}{0.38} & \mstd{0.88}{0.14} \\
UGATIT & \mstd{19.82}{4.00} & \mstd{0.59}{0.12} & \mstd{233.88}{39.94} & \mstd{0.62}{0.32} & \mstd{0.52}{0.30} & \mstd{42.43}{50.34} & \mstd{0.67}{0.35} & \mstd{0.91}{0.12} \\
CUT & \mstd{19.39}{3.75} & \mstd{0.56}{0.12} & \mstd{237.12}{37.46} & \mstd{0.58}{0.35} & \mstd{0.48}{0.33} & \mstd{48.14}{53.26} & \mstd{0.63}{0.37} & \mstd{0.91}{0.12} \\
FastCUT & \mstd{18.92}{4.35} & \mstd{0.56}{0.12} & \mstd{234.02}{36.81} & \mstd{0.42}{0.36} & \mstd{0.33}{0.32} & \mstd{69.80}{70.08} & \mstd{0.43}{0.38} & \bmstd{0.95}{0.09} \\
ACL-GAN & \mstd{14.80}{1.86} & \mstd{0.50}{0.12} & \mstd{295.66}{35.22} & \mstd{0.47}{0.28} & \mstd{0.35}{0.24} & \mstd{47.23}{36.76} & \mstd{0.72}{0.20} & \mstd{0.71}{0.19} \\
NiceGAN & \mstd{19.12}{4.08} & \mstd{0.57}{0.12} & \mstd{237.10}{40.66} & \mstd{0.52}{0.36} & \mstd{0.42}{0.32} & \mstd{64.83}{63.14} & \mstd{0.55}{0.39} & \mstd{0.92}{0.12} \\
AttentionGAN & \mstd{19.37}{4.15} & \mstd{0.57}{0.12} & \mstd{235.14}{39.91} & \mstd{0.60}{0.33} & \mstd{0.50}{0.31} & \mstd{46.31}{52.21} & \mstd{0.66}{0.36} & \mstd{0.91}{0.12} \\
QSGAN & \mstd{19.41}{3.76} & \mstd{0.56}{0.12} & \mstd{236.87}{37.72} & \mstd{0.57}{0.35} & \mstd{0.47}{0.33} & \mstd{48.29}{53.88} & \mstd{0.61}{0.38} & \mstd{0.91}{0.12} \\
DecentGAN & \mstd{19.64}{3.62} & \mstd{0.55}{0.12} & \mstd{239.42}{37.57} & \mstd{0.64}{0.31} & \mstd{0.54}{0.30} & \mstd{40.42}{50.99} & \mstd{0.70}{0.34} & \mstd{0.91}{0.12} \\
VQ-I2I & \mstd{12.56}{1.93} & \mstd{0.13}{0.07} & \mstd{303.43}{22.71} & \mstd{0.25}{0.25} & \mstd{0.17}{0.18} & \mstd{78.41}{55.93} & \mstd{0.27}{0.26} & \mstd{0.74}{0.23} \\
UVCGAN & \mstd{19.30}{3.83} & \mstd{0.57}{0.12} & \mstd{237.25}{37.65} & \mstd{0.58}{0.33} & \mstd{0.48}{0.31} & \mstd{49.48}{55.26} & \mstd{0.65}{0.36} & \mstd{0.88}{0.14} \\
SANTA & \mstd{18.98}{3.70} & \mstd{0.52}{0.12} & \mstd{247.15}{39.04} & \mstd{0.60}{0.32} & \mstd{0.50}{0.30} & \mstd{43.84}{52.89} & \mstd{0.65}{0.35} & \mstd{0.90}{0.14} \\
UNSB & \mstd{18.88}{3.86} & \mstd{0.54}{0.12} & \mstd{239.71}{38.87} & \mstd{0.62}{0.33} & \mstd{0.52}{0.31} & \mstd{49.67}{54.53} & \mstd{0.64}{0.36} & \mstd{0.92}{0.11} \\
StegoGAN & \mstd{19.41}{4.21} & \mstd{0.57}{0.12} & \mstd{234.33}{40.66} & \mstd{0.62}{0.31} & \mstd{0.52}{0.30} & \mstd{42.36}{49.33} & \mstd{0.69}{0.35} & \mstd{0.90}{0.13} \\
\hline
Pix2Pix & \mstd{18.78}{3.68} & \mstd{0.48}{0.14} & \mstd{250.81}{45.06} & \mstd{0.58}{0.36} & \mstd{0.49}{0.33} & \mstd{53.29}{65.45} & \mstd{0.61}{0.38} & \mstd{0.93}{0.09} \\
PyramidP2P & \mstd{21.37}{3.13} & \mstd{0.60}{0.10} & \mstd{224.87}{29.90} & \mstd{0.69}{0.31} & \mstd{0.59}{0.30} & \mstd{32.50}{46.53} & \mstd{0.70}{0.32} & \mstd{0.94}{0.08} \\
AdaptiveNCE & \mstd{21.38}{3.15} & \mstd{0.61}{0.11} & \mstd{222.95}{34.09} & \mstd{0.69}{0.30} & \mstd{0.59}{0.30} & \mstd{32.11}{45.42} & \bmstd{0.74}{0.31} & \mstd{0.94}{0.08} \\
VQ-I2I-Paired & \mstd{12.54}{1.96} & \mstd{0.13}{0.04} & \mstd{305.22}{14.13} & \mstd{0.15}{0.19} & \mstd{0.09}{0.13} & \mstd{105.39}{65.62} & \mstd{0.16}{0.21} & \mstd{0.84}{0.18} \\
CycleDiffusion & \mstd{15.73}{1.74} & \mstd{0.44}{0.10} & \mstd{300.17}{36.64} & \mstd{0.38}{0.28} & \mstd{0.28}{0.22} & \mstd{74.22}{60.60} & \mstd{0.56}{0.28} & \mstd{0.78}{0.14} \\
BBDM & \mstd{20.03}{3.74} & \mstd{0.57}{0.12} & \mstd{224.63}{38.15} & \mstd{0.63}{0.32} & \mstd{0.53}{0.31} & \mstd{40.33}{52.87} & \mstd{0.63}{0.34} & \mstd{0.94}{0.09} \\
LBBDM-f4 & \mstd{18.45}{2.42} & \mstd{0.45}{0.09} & \mstd{257.44}{37.28} & \mstd{0.56}{0.37} & \mstd{0.47}{0.34} & \mstd{52.74}{62.85} & \mstd{0.57}{0.39} & \mstd{0.92}{0.10} \\
LBBDM-f16 & \mstd{16.20}{1.92} & \mstd{0.26}{0.10} & \mstd{277.92}{37.64} & \mstd{0.50}{0.38} & \mstd{0.42}{0.34} & \mstd{63.81}{79.42} & \mstd{0.53}{0.40} & \mstd{0.90}{0.13} \\
Dual\_BranchPix2Pix & \mstd{20.99}{3.09} & \mstd{0.56}{0.11} & \mstd{227.78}{34.13} & \mstd{0.69}{0.31} & \mstd{0.60}{0.30} & \mstd{32.65}{46.70} & \mstd{0.72}{0.33} & \mstd{0.93}{0.09} \\
MCSStain &
\mstd{20.94}{2.98} &
\mstd{0.56}{0.10} &
\mstd{228.50}{34.41} &
\mstd{0.68}{0.31} &
\mstd{0.59}{0.30} &
\mstd{31.71}{46.06} &
\mstd{0.69}{0.32} &
\mstd{0.94}{0.08} \\
M2PL-GAN &
\mstd{20.58}{3.18} &
\mstd{0.57}{0.10} &
\mstd{231.5}{34.12} &
\mstd{0.67}{0.31} &
\mstd{0.57}{0.30} &
\mstd{34.74}{46.27} &
\mstd{0.71}{0.32} &
\mstd{0.92}{0.10} \\
ImplicitStainer &
\bmstd{22.25}{3.16} &
\bmstd{0.64}{0.10} &
\bmstd{219.20}{34.56} &
\bmstd{0.71}{0.31} &
\bmstd{0.61}{0.30} &
\bmstd{30.20}{45.05} &
\mstd{0.73}{0.33} &
\mstd{0.94}{0.09} \\
\bottomrule
\end{tabular}}
\caption{Mean $\pm$ standard deviation of PSNR, SSIM, MSE, Dice, Jaccard, HD, TPR, and TNR. Means for PSNR, MSE, and HD and all standard deviations are rounded to two decimal places for CK818 Dataset.}
\label{tab:metrics_ck818_mean_stddev}
\end{table*}

\begin{table*}[!htb]
\centering
\scriptsize
\setlength{\tabcolsep}{3pt}
\renewcommand{\arraystretch}{1.1}

\resizebox{\textwidth}{!}{%
\begin{tabular}{lcccccccc}
\toprule
Method & PSNR & SSIM & MSE & Dice & IoU & HD & TPR & TNR \\
\midrule
CycleGAN & \mstd{19.42}{3.75} & \mstd{0.55}{0.13} & \mstd{219.61}{40.39} & \mstd{0.54}{0.20} & \mstd{0.40}{0.18} & \mstd{21.51}{15.12} & \mstd{0.54}{0.20} & \mstd{0.99}{0.01} \\
UNIT & \mstd{17.28}{3.10} & \mstd{0.50}{0.12} & \mstd{249.07}{32.27} & \mstd{0.46}{0.16} & \mstd{0.32}{0.14} & \mstd{26.45}{16.79} & \mstd{0.48}{0.19} & \mstd{0.97}{0.02} \\
UGATIT & \mstd{19.21}{3.79} & \mstd{0.55}{0.13} & \mstd{218.57}{40.92} & \mstd{0.49}{0.20} & \mstd{0.35}{0.18} & \mstd{28.46}{19.41} & \mstd{0.45}{0.21} & \mstd{0.99}{0.01} \\
CUT & \mstd{19.50}{3.58} & \mstd{0.54}{0.13} & \mstd{221.37}{39.17} & \mstd{0.51}{0.21} & \mstd{0.37}{0.19} & \mstd{23.03}{15.68} & \mstd{0.50}{0.20} & \mstd{0.99}{0.01} \\
FastCUT & \mstd{19.50}{3.76} & \mstd{0.54}{0.13} & \mstd{221.46}{36.43} & \mstd{0.50}{0.20} & \mstd{0.36}{0.18} & \mstd{23.53}{15.49} & \mstd{0.48}{0.19} & \mstd{0.99}{0.01} \\
ACL-GAN & \mstd{16.10}{3.43} & \mstd{0.47}{0.12} & \mstd{262.51}{35.13} & \mstd{0.36}{0.15} & \mstd{0.23}{0.12} & \mstd{27.63}{13.17} & \mstd{0.38}{0.16} & \mstd{0.97}{0.02} \\
NiceGAN & \mstd{18.94}{3.70} & \mstd{0.55}{0.13} & \mstd{226.43}{42.97} & \mstd{0.49}{0.21} & \mstd{0.35}{0.18} & \mstd{24.52}{16.29} & \mstd{0.48}{0.22} & \mstd{0.99}{0.01} \\
AttentionGAN & \mstd{19.36}{3.75} & \mstd{0.55}{0.13} & \mstd{219.33}{39.89} & \mstd{0.54}{0.20} & \mstd{0.39}{0.19} & \mstd{22.06}{15.35} & \mstd{0.53}{0.21} & \mstd{0.99}{0.01} \\
QSGAN & \mstd{19.39}{3.57} & \mstd{0.54}{0.13} & \mstd{222.11}{39.12} & \mstd{0.50}{0.20} & \mstd{0.36}{0.18} & \mstd{22.98}{15.08} & \mstd{0.49}{0.19} & \mstd{0.98}{0.01} \\
DecentGAN & \mstd{19.20}{3.51} & \mstd{0.52}{0.13} & \mstd{226.98}{40.80} & \mstd{0.50}{0.19} & \mstd{0.36}{0.17} & \mstd{23.15}{15.24} & \mstd{0.51}{0.19} & \mstd{0.98}{0.01} \\
VQ-I2I & \mstd{18.22}{2.74} & \mstd{0.34}{0.11} & \mstd{271.45}{48.24} & \mstd{0.39}{0.18} & \mstd{0.26}{0.14} & \mstd{33.40}{19.17} & \mstd{0.40}{0.20} & \mstd{0.98}{0.02} \\

UVCGAN & \mstd{19.47}{3.64} & \mstd{0.55}{0.13} & \mstd{222.93}{40.12} & \mstd{0.53}{0.20} & \mstd{0.39}{0.18} & \mstd{21.74}{14.57} & \mstd{0.56}{0.20} & \mstd{0.98}{0.01} \\
SANTA & \mstd{18.90}{3.54} & \mstd{0.52}{0.13} & \mstd{227.70}{40.39} & \mstd{0.46}{0.20} & \mstd{0.32}{0.18} &  \mstd{22.25}{12.51}  & \mstd{0.46}{0.22} & \mstd{0.98}{0.01} \\
StegoGAN & \mstd{19.38}{3.79} & \mstd{0.55}{0.13} & \mstd{220.74}{40.81} & \mstd{0.53}{0.20} & \mstd{0.39}{0.18} & \mstd{22.37}{15.73} & \mstd{0.52}{0.21} & \mstd{0.99}{0.01} \\
UNSB & \mstd{19.31}{3.48} & \mstd{0.53}{0.13} & \mstd{225.73}{38.00} & \mstd{0.49}{0.20} & \mstd{0.35}{0.18} & \mstd{23.95}{16.24} & \mstd{0.47}{0.21} & \mstd{0.99}{0.01} \\
\midrule
Pix2Pix & \mstd{19.01}{3.44} & \mstd{0.46}{0.12} & \mstd{234.63}{36.47} & \mstd{0.54}{0.19} & \mstd{0.39}{0.17} & \mstd{23.86}{16.66} & \mstd{0.56}{0.19} & \mstd{0.98}{0.01} \\
PyramidP2P & \mstd{20.09}{3.50} & \mstd{0.54}{0.13} & \mstd{215.69}{38.46} & \mstd{0.59}{0.19} & \mstd{0.45}{0.18} & \mstd{18.59}{13.71} & \mstd{0.60}{0.18} & \mstd{0.99}{0.01} \\
AdaptiveNCE & \mstd{20.45}{3.58} & \mstd{0.57}{0.13} & \mstd{213.23}{38.52} & \mstd{0.59}{0.20} & \mstd{0.45}{0.19} & \bmstd{17.58}{13.48} & \mstd{0.60}{0.19} & \mstd{0.99}{0.01} \\
VQ-I2I-Paired & \mstd{16.39}{4.67} & \mstd{0.30}{0.12} & \mstd{253.03}{40.85} & \mstd{0.29}{0.20} & \mstd{0.19}{0.14} & \mstd{70.38}{51.77} & \mstd{0.25}{0.20} & \mstd{0.99}{0.01} \\
CycleDiffusion & \mstd{15.31}{1.63} & \mstd{0.41}{0.10} & \mstd{312.48}{30.57} & \mstd{0.40}{0.17} & \mstd{0.26}{0.13} & \mstd{31.69}{17.43} & \mstd{0.51}{0.19} & \mstd{0.96}{0.02} \\
BBDM & \mstd{19.80}{3.76} & \mstd{0.55}{0.13} & \mstd{212.80}{40.94} & \mstd{0.58}{0.20} & \mstd{0.44}{0.18} & \mstd{19.12}{13.55} & \mstd{0.61}{0.19} & \mstd{0.98}{0.02} \\
LBBDM-f4 & \mstd{18.46}{3.14} & \mstd{0.44}{0.12} & \mstd{244.84}{41.18} & \mstd{0.39}{0.16} & \mstd{0.26}{0.13} & \mstd{32.83}{17.12} & \mstd{0.38}{0.16} & \mstd{0.98}{0.02} \\
LBBDM-f16 & \mstd{16.94}{2.70} & \mstd{0.29}{0.11} & \mstd{266.40}{40.21} & \mstd{0.26}{0.12} & \mstd{0.15}{0.08} & \mstd{47.14}{21.51} & \mstd{0.24}{0.13} & \mstd{0.98}{0.01} \\
Dual\_BranchPix2Pix & \mstd{20.21}{3.41} & \mstd{0.52}{0.13} & \mstd{219.88}{39.32} & \mstd{0.60}{0.19} & \mstd{0.44}{0.18} & \mstd{19.75}{13.31} & \mstd{0.61}{0.20} & \mstd{0.99}{0.01} \\
MCSStain & \mstd{20.02}{3.41} & \mstd{0.51}{0.12} & \mstd{223.91}{37.74} & \mstd{0.60}{0.19} & \mstd{0.45}{0.18} & \bmstd{17.58}{12.92} & \bmstd{0.62}{0.19} & \mstd{0.98}{0.01} \\
PPT & \mstd{19.83}{3.35} & \mstd{0.49}{0.12} & \mstd{230.51}{40.12} & \mstd{0.49}{0.19} & \mstd{0.34}{0.17} & \mstd{27.23}{17.58} & \mstd{0.44}{0.20} & \mstd{0.99}{0.01} \\
M2PL & \mstd{20.29}{3.55} & \mstd{0.56}{0.13} & \mstd{216.70}{38.62} & \mstd{0.57}{0.19} & \mstd{0.43}{0.18} & \mstd{18.59}{13.36} & \mstd{0.58}{0.20} & \mstd{0.99}{0.01} \\
\bf ImplicitStainer(Ours) &  \bmstd{21.28}{3.53} & \bmstd{0.59}{0.13} & \bmstd{211.91}{42.60} & \bmstd{0.61}{0.19} & \bmstd{0.46}{0.18} & \mstd{17.78}{14.62} & \mstd{0.60}{0.19} & \mstd{0.99}{0.01} \\
\bottomrule
\end{tabular}%
}
\caption{Mean $\pm$ standard deviation of PSNR, SSIM, MSE, Dice, Jaccard, HD, TPR, and TNR. Means for PSNR, MSE, and HD and all standard deviations are rounded to two decimal places for IHC CD3  Dataset.}
\label{tab:metrics_cd3_mean_stddev}
\end{table*}

\begin{table*}[t]
\centering
\scriptsize
\setlength{\tabcolsep}{2.5pt}
\renewcommand{\arraystretch}{1.1}
\resizebox{\textwidth}{!}{%
\begin{tabular}{l|ccc||ccccccccc}
\toprule
& \multicolumn{3}{c||}{ } & \multicolumn{3}{c|}{\it DAPI} & \multicolumn{3}{c|}{\it CD3} &  \multicolumn{3}{c}{\it PanCK}\\  
Method &
PSNR &
SSIM &
MSE &
Dice &
Jaccard &
HD &
Dice &
Jaccard &
HD &
Dice &
Jaccard &
HD \\
\midrule

CycleGAN &
\mstd{21.32}{1.95} &
\mstd{0.55}{0.08} &
\mstd{149.93}{28.46} &
\mstd{0.28}{0.29} &
\mstd{0.20}{0.23} &
\mstd{101.51}{86.99} &
\mstd{0.03}{0.05} &
\mstd{0.02}{0.03} &
\mstd{104.85}{70.21} &
\mstd{0.05}{0.06} &
\mstd{0.02}{0.03} &
\mstd{64.56}{35.98} \\
UNIT &
\mstd{19.54}{3.36} &
\mstd{0.65}{0.15} &
\mstd{136.39}{46.59} &
\mstd{0.45}{0.37} &
\mstd{0.37}{0.33} &
\mstd{61.55}{70.99} &
\mstd{0.18}{0.23} &
\mstd{0.12}{0.17} &
\mstd{108.58}{81.11} &
\mstd{0.68}{0.27} &
\mstd{0.58}{0.28} &
\mstd{25.84}{34.55} \\
UGATIT &
\mstd{25.19}{3.77} &
\mstd{0.74}{0.13} &
\mstd{125.97}{46.42} &
\mstd{0.44}{0.36} &
\mstd{0.36}{0.32} &
\mstd{68.97}{72.79} &
\mstd{0.02}{0.04} &
\mstd{0.01}{0.02} &
\mstd{108.86}{78.23} &
\mstd{0.54}{0.30} &
\mstd{0.43}{0.28} &
\mstd{33.14}{30.78} \\
CUT &
\mstd{21.37}{2.17} &
\mstd{0.51}{0.07} &
\mstd{165.37}{28.34} &
\mstd{0.08}{0.15} &
\mstd{0.05}{0.11} &
\mstd{110.81}{84.66} &
\mstd{0.03}{0.05} &
\mstd{0.01}{0.03} &
\mstd{108.28}{85.76} &
\mstd{0.07}{0.09} &
\mstd{0.04}{0.05} &
\mstd{60.16}{33.83} \\
FastCUT &
\mstd{26.01}{3.67} &
\mstd{0.75}{0.11} &
\mstd{94.44}{37.81} &
\mstd{0.51}{0.34} &
\mstd{0.41}{0.31} &
\mstd{64.18}{74.07} &
\mstd{0.19}{0.24} &
\mstd{0.13}{0.18} &
\mstd{88.21}{81.01} &
\mstd{0.77}{0.18} &
\mstd{0.66}{0.21} &
\mstd{18.03}{25.48} \\
NiceGAN &
\mstd{24.36}{3.42} &
\mstd{0.54}{0.12} &
\mstd{112.20}{46.36} &
\mstd{0.00}{0.00} &
\mstd{0.00}{0.00} &
\mstd{0.00}{0.00} &
\mstd{0.00}{0.00} &
\mstd{0.00}{0.00} &
\mstd{0.00}{0.00} &
\mstd{0.00}{0.00} &
\mstd{0.00}{0.00} &
\mstd{0.00}{0.00} \\

AttentionGAN &
\mstd{25.45}{3.81} &
\mstd{0.76}{0.12} &
\mstd{91.15}{43.38} &
\mstd{0.49}{0.34} &
\mstd{0.39}{0.31} &
\mstd{65.80}{72.20} &
\mstd{0.01}{0.04} &
\mstd{0.01}{0.02} &
\mstd{100.37}{82.56} &
\mstd{0.72}{0.19} &
\mstd{0.59}{0.22} &
\mstd{26.06}{29.84} \\
QSGAN &
\mstd{27.31}{3.99} &
\mstd{0.82}{0.10} &
\mstd{73.76}{38.40} &
\mstd{0.54}{0.35} &
\mstd{0.44}{0.32} &
\mstd{50.62}{65.55} &
\mstd{0.26}{0.27} &
\mstd{0.18}{0.21} &
\mstd{79.72}{72.89} &
\mstd{0.81}{0.17} &
\mstd{0.71}{0.21} &
\mstd{13.70}{23.31} \\
DecentGAN &
\mstd{23.44}{2.94} &
\mstd{0.65}{0.13} &
\mstd{124.44}{52.47} &
\mstd{0.20}{0.22} &
\mstd{0.13}{0.16} &
\mstd{98.06}{75.68} &
\mstd{0.16}{0.22} &
\mstd{0.11}{0.15} &
\mstd{93.46}{82.07} &
\mstd{0.51}{0.24} &
\mstd{0.38}{0.22} &
\mstd{41.44}{35.10} \\

VQ-I2I &
\mstd{22.88}{2.38} &
\mstd{0.58}{0.08} &
\mstd{125.32}{29.49} &
\mstd{0.19}{0.19} &
\mstd{0.12}{0.13} &
\mstd{95.64}{75.41} &
\mstd{0.06}{0.11} &
\mstd{0.04}{0.07} &
\mstd{104.59}{75.48} &
\mstd{0.26}{0.13} &
\mstd{0.15}{0.09} &
\mstd{55.67}{34.59} \\

UNSB &
\mstd{21.90}{2.40} &
\mstd{0.54}{0.09} &
\mstd{157.75}{29.65} &
\mstd{0.17}{0.22} &
\mstd{0.11}{0.16} &
\mstd{103.49}{79.29} &
\mstd{0.03}{0.06} &
\mstd{0.02}{0.04} &
\mstd{120.74}{79.19} &
\mstd{0.22}{0.15} &
\mstd{0.13}{0.10} &
\mstd{58.74}{35.32} \\
StegoGAN &
\mstd{22.11}{2.11} &
\mstd{0.59}{0.08} &
\mstd{134.48}{26.80} &
\mstd{0.30}{0.31} &
\mstd{0.22}{0.24} &
\mstd{101.48}{87.35} &
\mstd{0.02}{0.05} &
\mstd{0.01}{0.03} &
\mstd{108.23}{72.49} &
\mstd{0.02}{0.05} &
\mstd{0.01}{0.03} &
\mstd{80.50}{50.99} \\

\hline
Pix2Pix &
\mstd{28.61}{4.54} &
\mstd{0.84}{0.10} &
\mstd{64.01}{39.01} &
\mstd{0.54}{0.36} &
\mstd{0.45}{0.33} &
\mstd{52.93}{68.13} &
\mstd{0.24}{0.26} &
\mstd{0.17}{0.20} &
\mstd{73.68}{68.98} &
\mstd{0.80}{0.21} &
\mstd{0.71}{0.23} &
\mstd{13.84}{24.44} \\
PyramidP2P &
\mstd{29.26}{4.30} &
\mstd{0.85}{0.09} &
\mstd{61.28}{37.40} &
\mstd{0.57}{0.35} &
\mstd{0.48}{0.33} &
\mstd{49.72}{65.07} &
\mstd{0.32}{0.31} &
\mstd{0.23}{0.25} &
\mstd{68.23}{70.05} &
\mstd{0.82}{0.18} &
\mstd{0.73}{0.21} &
\mstd{12.92}{23.39} \\
AdaptiveNCE &
\mstd{27.11}{3.48} &
\mstd{0.45}{0.10} &
\mstd{129.83}{27.61} &
\mstd{0.61}{0.38} &
\mstd{0.53}{0.36} &
\mstd{48.64}{77.23} &
\mstd{0.10}{0.12} &
\mstd{0.06}{0.08} &
\mstd{110.95}{79.49} &
\mstd{0.81}{0.13} &
\mstd{0.70}{0.17} &
\mstd{13.72}{20.07} \\
VQ-I2I-Paired &
\mstd{26.04}{3.31} &
\mstd{0.78}{0.09} &
\mstd{81.53}{35.18} &
\mstd{0.51}{0.35} &
\mstd{0.41}{0.31} &
\mstd{62.90}{74.31} &
\mstd{0.26}{0.24} &
\mstd{0.17}{0.18} &
\mstd{88.45}{71.77} &
\mstd{0.74}{0.15} &
\mstd{0.61}{0.17} &
\mstd{17.43}{22.88} \\
Dual\_BranchPix2Pix &
\mstd{28.08}{3.70} &
\mstd{0.82}{0.09} &
\mstd{68.32}{36.44} &
\mstd{0.52}{0.35} &
\mstd{0.42}{0.32} &
\mstd{57.71}{67.36} &
\mstd{0.21}{0.25} &
\mstd{0.14}{0.19} &
\mstd{80.61}{76.94} &
\mstd{0.78}{0.22} &
\mstd{0.68}{0.24} &
\mstd{17.42}{27.98} \\
PPT &
\mstd{27.05}{4.02} &
\mstd{0.80}{0.11} &
\mstd{77.80}{42.45} &
\mstd{0.50}{0.34} &
\mstd{0.34}{0.30} &
\mstd{60.34}{67.33} &
\mstd{0.21}{0.23} &
\mstd{0.14}{0.16} &
\mstd{85.04}{73.76} &
\mstd{0.77}{0.17} &
\mstd{0.65}{0.19} &
\mstd{15.74}{22.52} \\
MCSStain &
\mstd{30.87}{5.04} &
\bmstd{0.88}{0.09} &
\mstd{55.82}{36.43} &
\mstd{0.61}{0.36} &
\mstd{0.53}{0.35} &
\mstd{49.87}{60.42} &
\mstd{0.36}{0.35} &
\mstd{0.28}{0.29} &
\mstd{67.64}{71.60} &
\mstd{0.84}{0.18} &
\mstd{0.75}{0.21} &
\mstd{13.86}{22.16} \\
M2PL &
\mstd{28.37}{4.56} &
\mstd{0.84}{0.10} &
\mstd{63.72}{39.06} &
\mstd{0.56}{0.36} &
\mstd{0.47}{0.33} &
\mstd{52.41}{69.20} &
\mstd{0.27}{0.26} &
\mstd{0.18}{0.20} &
\mstd{78.57}{68.80} &
\mstd{0.83}{0.17} &
\mstd{0.74}{0.20} &
\mstd{12.83}{21.73} \\

ImplicitStainer &
\bmstd{32.26}{4.61} &
\mstd{0.85}{0.13} &
\bmstd{55.39}{40.15} &
\bmstd{0.64}{0.35} &
\bmstd{0.55}{0.33} &
\bmstd{43.78}{65.47} &
\bmstd{0.50}{0.36} &
\bmstd{0.41}{0.32} &
\bmstd{58.30}{75.12} &
\bmstd{0.85}{0.16} &
\bmstd{0.77}{0.19} &
\bmstd{12.58}{23.68} \\
\bottomrule
\end{tabular}}
\caption{Mean $\pm$ standard deviation of PSNR, SSIM, MSE, Dice, Jaccard and Hausdorff for each of the stain in HEMIT dataset.}
\label{tab:hemit_metrics_mean_std}
\end{table*}

\begin{table*}[!htb]
\scalebox{0.95}{
\setlength{\tabcolsep}{4pt}
    \centering
    \begin{tabular}{c||c|c|c||c|c|c|c||c|c|c|c|c}
    & \multicolumn{3}{c||}{\bf Texture Metrics } &  \multicolumn{4}{c||}{\bf Distribution Metrics} & \multicolumn{5}{c}{\bf Segmentation Metrics} \\
    I2I Methods & PSNR $\uparrow$ & SSIM $\uparrow$ & MSE $\downarrow$& FID $\downarrow$& KID $\downarrow$ & Prec $\uparrow$ & Rec $\uparrow$  & Dice $\uparrow$ & IoU $\uparrow$ & HD$\downarrow$ & TPR$\uparrow$ & TNR$\uparrow$ \\
    \hline
    \hline
     {\name }   & \bf 21.28	& \bf 0.5937 & \bf 211.91	& 33.79	& 0.0334	& 0.7622 & \bf 0.6709 & \bf 0.6081 & \bf 0.4617 &	\bf  17.77 &	 0.5980 & \bf 0.9877 \\
      {\name} \{Conv\}   &  20.79	& 0.5686 & 218.85 & \bf 27.33	& \bf 0.0274	& \bf 0.8331	& 0.6329& 0.5726 & 0.4256 & 18.65 &	\bf 0.5989	& 0.9831 \\
     {\name }  \{Swin\} & 20.49 & 0.5667&  237.49& 47.06&  0.0538 & 0.7527 & 0.5457 & 0.5487	& 0.4007	& 21.47 & 0.5545& 0.9848\\
    \end{tabular}}
   \caption{\textbf{Backbone Ablation on the CD3 IHC Dataset.} Quantitative comparison of a convolution-only backbone, a Swin Transformer-only backbone, and the proposed hybrid convolution--Swin architecture. The hybrid model achieves the strongest overall performance across most image-fidelity and segmentation metrics, indicating that combining local convolutional features with global transformer-based context provides a more effective representation than either backbone alone for this task.
}
    \label{tab:quantiative_metrics_ablation_CD3_backbone}
    \vspace{-1em}
\end{table*}

\begin{table*}[!htb]
\scalebox{0.90}{
\setlength{\tabcolsep}{5pt}
    \begin{tabular}{p{0.9cm}|c||c|c|c||c|c|c|c||c|c|c|c|c}
   & & \multicolumn{3}{c||}{\bf Texture Metrics } &  \multicolumn{4}{c||}{\bf Distribution Metrics} & \multicolumn{5}{c}{\bf Segmentation Metrics} \\
  dataset & ($\lambda_1$, $\lambda_2$, $\lambda_3$)& PSNR $\uparrow$ & SSIM $\uparrow$ & MSE $\downarrow$& FID $\downarrow$& KID $\downarrow$ & Prec $\uparrow$ & Rec $\uparrow$  & Dice $\uparrow$ & IoU $\uparrow$ & HD$\downarrow$ & TPR$\uparrow$ & TNR$\uparrow$ \\
    \hline 
     CK818 &   ($1.0$, $1.0$, $0$)            &   22.24	&  0.644	& 219.19 & 17.69 & 0.0107 &	0.9131 &	0.8410 & 0.7049	&  0.6149	&  30.19	& 0.7298	& 0.9351 \\
     CK818 &   ($0$, $0$, $0$)            &   21.33	&  0.6206	&  226.25 & 116.16 & 0.1125 & 0.1790 &	0.1764 & 0.6351	&  0.5366	&  44.99	& 0.6591	& 0.9252 \\
     \hline
     CD3 &    ($1.0$, $1.0$, $0$)              &  21.28	&  0.5937 & 211.91	& 33.79	& 0.0334	& 0.7622 & 0.6709 &  0.6081 & 0.4617 &	17.77 &	0.5980 & 0.9877 \\
     CD3 &   ($0$, $0$, $0$)            &   21.00	&  0.5931	&  212.5 & 101.61 & 0.1080 &	 0.1437 &	0.1788 & 0.5993	&  0.4521	&  20.50	& 0.5809	& 0.9890 \\
    \end{tabular}}
    \caption{\textbf{Ablation Experiments without Perceptual Loss.} Models are trained by setting the perceptual loss coefficients associated with different encoders to zero. Removing the perceptual loss leads to a significant degradation in distribution-based metrics, while the decline in texture and segmentation-based metrics is comparatively smaller. These results suggest that perceptual loss primarily contributes to improving visual realism and distributional alignment, whereas staining accuracy is governed more strongly by the underlying model architecture.}
    \label{tab:without_perceptual_loss}
\end{table*}

\begin{table*}[!thb]
\setlength{\tabcolsep}{2.0pt}
\scalebox{0.9}{
    \begin{tabular}{ccc||ccc||ccc||ccc|ccc|ccc}
   & & & \multicolumn{3}{c||}{\bf Texture Metrics } &  \multicolumn{3}{c||}{\bf Distribution Metrics} & \multicolumn{9}{c}{\bf Segmentation Metrics}\\ 
   & & & \multicolumn{3}{c||}{ } &  \multicolumn{3}{c||}{} & \multicolumn{3}{c|}{\it DAPI} & \multicolumn{3}{c|}{\it CD3} &  \multicolumn{3}{c}{\it PanCK}\\  
    $\lambda_1$ & $\lambda_2$ & $\lambda_3$ & PSNR $\uparrow$ & SSIM $\uparrow$ & MSE$\downarrow$& FID $\downarrow$ & Prec $\uparrow$ & Recall $\uparrow$&  Dice $\uparrow$ & IoU $\uparrow$ & HD $\downarrow$ &  Dice $\uparrow$ & IoU $\uparrow$ & HD $\downarrow$ &  Dice $\uparrow$ & IoU $\uparrow$ & HD $\downarrow$\\
    \hline
    \hline
    1 & 1 & 1 & 29.8192 & 0.8160 & 67.1531 & 24.9570 & \bf 0.8591 & 0.6211  & 0.6072 & 0.5114 & 52.3504  & 0.4064 & 0.3141 & 68.3411  & 0.8362 & 0.7461 & 12.8636 \\
    1 & 1 & 0 &  31.0629 & 0.7959 & 65.1271 & \bf 16.0185 & 0.8154 & \bf 0.7240 & \bf 0.6191 & \bf 0.5251 & 51.2701 & \bf 0.5336 & \bf 0.4409 & \bf 57.4083 & 0.8458 & 0.7585 & 13.9348 \\
    1 & 0 & 0 & 30.1243 & 0.7225 & 76.1369 & 50.4479 & 0.4823 & 0.4845  & 0.6003 & 0.5042 & 58.6871  & 0.5033 & 0.4071 & 56.4856  & 0.8198 & 0.7202 & 19.0277 \\
    10 & 10 & 0 & 30.7995 & 0.8082 & 65.7215 &  17.6840 & 0.7792 & 0.7125  & 0.6122 & 0.5170 & 52.9250  & 0.5335 & 0.4401 & \bf 55.6445  & 0.8438 & 0.7556 & 14.5861 \\
    10 & 0 & 0 & 29.0385 & 0.6490 & 87.6859 & 87.8634 & 0.2079 & 0.2264 & 0.5800 & 0.4816 & 67.9357 & 0.4948 & 0.3992 & 58.0852 & 0.7689 & 0.6501 & 27.1301 \\
    100 & 100 & 0 & 30.7117 & 0.8052 & 66.6226 & 18.0539 & 0.7872 & 0.7006 & 0.6112 & 0.5163 & 53.4505 & 0.5309 & 0.4376 & 56.0026 & 0.8439 & 0.7553 & 14.5175 \\
    100 & 0 & 0 & 28.4422 & 0.6140 & 93.9633 & 109.1401 & 0.1305 & 0.1526 & 0.5736 & 0.4737 & 69.3155 & 0.4837 & 0.3887 & 60.9142 & 0.7245 & 0.5940 & 31.6193 \\
    0 & 0 & 0 & \bf 32.7363 & \bf 0.8699 & \bf 52.4190 & 54.3857 & 0.4991 & 0.3322 & 0.6037 & 0.5118 & \bf 47.2539 & \bf 0.5360 & \bf 0.4428 & 64.0330 & \bf 0.8516 & \bf 0.7667 & 12.8915 \\
0 & 0.1 & 0 & 31.4949 & 0.8269 & 61.8438 & 17.9795 & 0.8302 & 0.7095 & 0.6151 & 0.5214 & 49.1322 & 0.5000 & 0.4093 & 66.0137 & 0.8486 & 0.7639 & 12.4544 \\
0 & 1 & 0 & 30.9365 & 0.7884 & 67.3255 & 19.1162 & 0.8006 & 0.6981 & 0.6132 & 0.5199 & 50.9185 & 0.4728 & 0.3868 & 67.3217 & 0.8397 & 0.7528 & 12.9627 \\
0.1 & 0.1 & 0.1 & 29.7720 & 0.7992 & 67.3035 & 24.2715 & 0.8235 & 0.5973 & 0.6035 & 0.5068 & 52.7743 & 0.3906 & 0.2948 & 75.2250 & 0.8361 & 0.7460 & 12.9253 \\
0.1 & 0 & 0 & 31.5545 & 0.8261 & 61.2673 & 19.7809 & 0.8009 & 0.6311 & 0.6145 & 0.5183 & 48.4796 & 0.5327 & 0.4361 & 55.0739 & 0.8509 & 0.7658 & \bf 12.0745 \\
0.1 & 0.1 & 0 & 31.5519 & 0.8238 & 61.1427 & \bf 14.1418 & \bf 0.8764 & \bf 0.7277 & \bf 0.6205 & \bf 0.5266 & \bf 47.2604 & 0.5268 & 0.4337 & 60.5318 & \bf 0.8522 & \bf 0.7680 & \bf 12.1802 \\
0 & 0 & 1 & 28.7761 & 0.7858 & 74.9733 & 34.5319 & 0.7376 & 0.4973 & 0.5761 & 0.4758 & 56.5582 & 0.3162 & 0.2313 & 78.1268 & 0.8273 & 0.7318 & 13.5782 \\
0 & 10 & 0 & 30.7287 & 0.7965 & 66.0181 & 19.4254 & 0.8067 & 0.6743 & 0.6076 & 0.5138 & 52.6649 & 0.4700 & 0.3829 & 67.5434 & 0.8359 & 0.7482 & 13.6057 \\
    \end{tabular}}
  \caption{\textbf{($\lambda$)-Value Ablation on the HEMIT Dataset.} Quantitative evaluation of different perceptual-loss configurations obtained by varying the loss weights (($\lambda_1,\lambda_2,\lambda_3$)) associated with AlexNet, VGG, and ResNet50 perceptual supervision, respectively. Results highlight the effect of both the number of perceptual encoders and their relative contributions to the overall objective. \textbf{Bold} denotes the top two performing configurations for each metric. For computational efficiency, \textit{Conv} variants of \name are used in these experiments due to their faster training speed.}
    \label{tab:quantiative_metrics_lambda}
\end{table*}

\begin{table*}[!thb]
\setlength{\tabcolsep}{2.0pt}
\scalebox{0.9}{
    \begin{tabular}{c||ccc||ccc||ccc|ccc|ccc}
     & \multicolumn{3}{c||}{\bf Texture Metrics } &  \multicolumn{3}{c||}{\bf Distribution Metrics} & \multicolumn{9}{c}{\bf Segmentation Metrics}\\ 
    & \multicolumn{3}{c||}{ } &  \multicolumn{3}{c||}{} & \multicolumn{3}{c|}{\it DAPI} & \multicolumn{3}{c|}{\it CD3} &  \multicolumn{3}{c}{\it PanCK}\\  
    channel & PSNR $\uparrow$ & SSIM $\uparrow$ & MSE$\downarrow$& FID $\downarrow$ & Prec $\uparrow$ & Recall $\uparrow$&  Dice $\uparrow$ & IoU $\uparrow$ & HD $\downarrow$ &  Dice $\uparrow$ & IoU $\uparrow$ & HD $\downarrow$ &  Dice $\uparrow$ & IoU $\uparrow$ & HD $\downarrow$\\
    \hline
   32 & 29.8152 &	0.8210 &	65.9443	& 25.6943	&0.8216	&0.6156	&0.6111	&0.5163	&50.8715	&0.3890	&0.2949	&73.3280	&0.8344	&0.7448	&13.7970 \\

   64 & \bf 30.6508 &	\bf 0.8393	& \bf 62.4183	& \bf20.3409	&\bf 0.8618	&\bf 0.7126	&\bf 0.6264	&\bf 0.5349	&\bf 46.6279	&\bf 0.4447	&\bf 0.3551	&\bf 63.9456	&\bf 0.8381	&\bf 0.7496	& 12.8846 \\

   256 & 29.8192&	0.8160 &	67.1531	&24.9570	&0.8591	&0.6211	&0.6072	&0.5114	&52.3504	&0.4064	&0.3141	&68.3411	&0.8362	&0.7461	& \bf 12.8636\\
    \hline
    \end{tabular}}
    \caption{\textbf{Ablation For Channel Dimension $C$ .} We evaluated three channel-dimension configurations—32, 64, and 256—as described in Figure~\ref{fig:pipeline_fig} and the Methods section. Channel dimension of 64 provides the best results. }
    \label{tab:quantiative_channel_dimension}
\end{table*}

\begin{table*}[!thb]
\setlength{\tabcolsep}{2pt}
\scalebox{0.9}{
    \begin{tabular}{c||ccc||ccc||ccc|ccc|ccc}
     & \multicolumn{3}{c||}{\bf Texture Metrics } &  \multicolumn{3}{c||}{\bf Distribution Metrics} & \multicolumn{9}{c}{\bf Segmentation Metrics}\\ 
    & \multicolumn{3}{c||}{ } &  \multicolumn{3}{c||}{} & \multicolumn{3}{c|}{\it DAPI} & \multicolumn{3}{c|}{\it CD3} &  \multicolumn{3}{c}{\it PanCK}\\  
    Window & PSNR $\uparrow$ & SSIM $\uparrow$ & MSE$\downarrow$& FID $\downarrow$ & Prec $\uparrow$ & Recall $\uparrow$&  Dice $\uparrow$ & IoU $\uparrow$ & HD $\downarrow$ &  Dice $\uparrow$ & IoU $\uparrow$ & HD $\downarrow$ &  Dice $\uparrow$ & IoU $\uparrow$ & HD $\downarrow$\\
    \hline
   3 & 29.4525&	0.8007	& 69.6850	&28.0398	& \bf 0.8535 &	0.5794&	0.6034	&0.5072	& 54.8674	& 0.3859	& 0.2921&	71.4931	&0.8310	&0.7392	&14.7747 \\

   5 & 29.3916	&0.7997	&70.0891&	26.9074	&0.8407	&0.5797	&0.6022	&0.5055	&54.6958	&0.3789	&0.2834	&72.6687	&0.8321	&0.7410	&14.3463 \\
   7 & \bf 29.6692 & \bf 0.8126 & \bf 67.5877 & \bf 25.52 & \ 0.8419 & \bf 0.6006 & \bf 0.6053 & \bf 0.5091 & \bf 52.6576 & \bf 0.3913 & \bf 0.2976 & \bf 69.3724 & \bf 0.8353 & \bf 0.7453 & \bf 13.7406\\
    \hline
    \end{tabular}}
  \caption{\textbf{Window Size Ablation ($w$).} Quantitative comparison of different neighborhood window sizes used in Eq.~6. Larger windows generally improve image-fidelity, distribution-based, and segmentation metrics, with $w=7$ achieving the strongest overall performance. However, these gains are accompanied by substantially higher memory and computational requirements. Given the relatively modest improvements in staining accuracy compared with the increased computational cost, we select $w=3$ as the default configuration for all subsequent experiments.
}
    \label{tab:quantiative_window_size_dimension}
\end{table*}

\begin{table*}[!htb]
\scalebox{0.90}{
\setlength{\tabcolsep}{5pt}
    \centering
    \begin{tabular}{p{0.1cm}|c||c|c|c||c|c|c|c||c|c|c|c|c}
   & & \multicolumn{3}{c||}{\bf Texture Metrics } &  \multicolumn{4}{c||}{\bf Distribution Metrics} & \multicolumn{5}{c}{\bf Segmentation Metrics} \\
   & I2I Methods & PSNR $\uparrow$ & SSIM $\uparrow$ & MSE $\downarrow$& FID $\downarrow$& KID $\downarrow$ & Prec $\uparrow$ & Rec $\uparrow$  & Dice $\uparrow$ & IoU $\uparrow$ & HD$\downarrow$ & TPR$\uparrow$ & TNR$\uparrow$ \\
    \hline
    &   {\name }   & 21.28	& 0.5937 & 211.91	& 33.79	& 0.0334	& 0.7622 & 0.6709 & 0.6081 &  0.4617 &	17.77 &	0.5980 & 0.9877 \\
    &  {{\sc ImplicitStainer-LR}}   &  20.49 & 0.4996 & 225.15 &  364.57 & 0.4317  & 0.0000  &  0.0004 & 0.5713 & 0.4231 & 20.80 &  0.5563 & 0.9875  \\
    \hline
    \end{tabular}}
    \caption{\textbf{Ablation Experiments with Low-Resolution Training and High-Resolution Inference on the CD3 Dataset.} In these experiments, {\name} refers to the model trained directly on high-resolution images ($256 \times 256$), whereas {\sc ImplicitStainer-LR} refers to the model trained on low-resolution images ($64 \times 64$) but evaluated at high resolution ($256 \times 256$) during inference using grid sampling. Both models were evaluated under identical high-resolution inference settings.}
    \label{tab:quantiative_metrics_ablations_lr_CD3}
\end{table*}

\begin{figure*}[!htb]
    \centering
    \includegraphics[trim={2.0cm 0.5cm 13.0cm 0.0cm}, clip=true, width=0.8\linewidth]{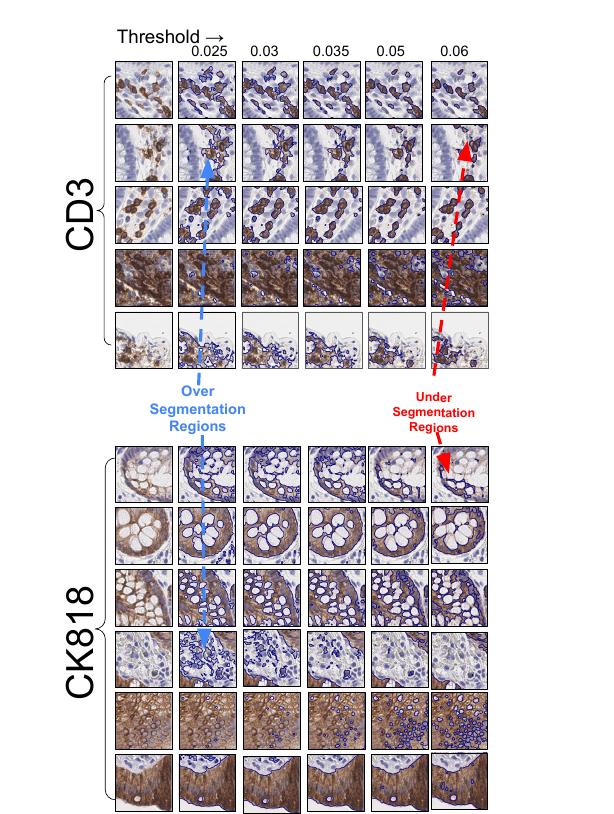}
    \caption{\textbf{Threshold Senstivity for CK818 and CD3.} These masks are shown as outlines for different threshold values of the DAB channel. As illustrated in the images, the segmentation results vary considerably with the chosen threshold. The image on the left, corresponding to a lower threshold, exhibits over-segmentation by incorrectly capturing tissue regions that are not positively stained. In contrast, the image on the right, corresponding to a higher threshold, demonstrates under-segmentation, where genuinely stained regions are missed. }
     \label{fig:Threshold_senstivity}
\end{figure*}

\begin{figure}[!htb]
    \centering
    \includegraphics[trim={0.0cm 7.25cm 18.2cm 0.0cm}, clip=true, width=1.0\linewidth]{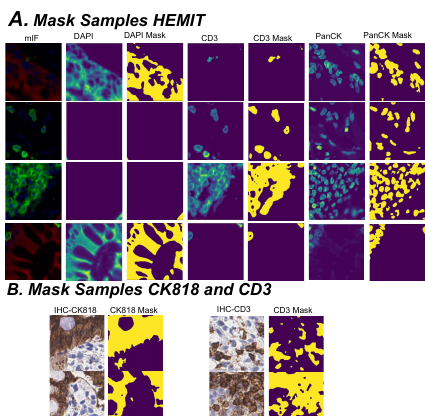}
     \caption{\textbf{Threshold Mask Samples.} (A) Binary masks of immunofluorescent antibody stains alongside their corresponding ground-truth images from HEMIT \cite{bian2024hemit}. (B) Binary masks extracted from the DAB channel for CK8/18 and CD3 stains. These examples illustrate that the proposed thresholding-based approach effectively identifies pixels corresponding to positive IHC and mIF signals.}
     \label{fig:qualitative_results_threshold}
\end{figure}

\begin{figure}[!htb]
    \centering
    \includegraphics[trim={1.3cm 7.25cm 15.4cm 0cm}, clip=true, width=1.0\linewidth]{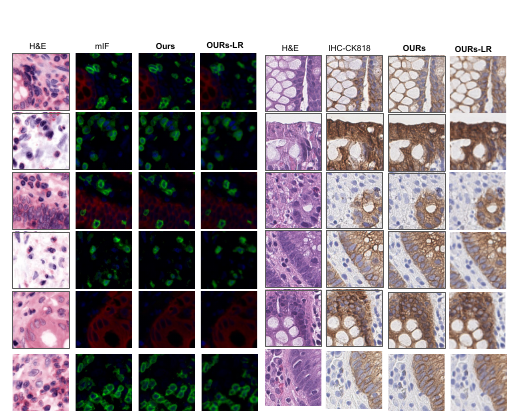}
     \caption{\textbf{High-Resolution Inference from Low-Resolution Training}. Qualitative results when \name is trained on low-resolution images but evaluated at high resolution using denser grid sampling within the implicit modeling block. {\sc Ours-LR} denotes the model trained on low-resolution images, while inference for all compared methods is consistently performed at high resolution.}
     \label{fig:low_resolution_inference}
\end{figure}

\begin{figure*}[!htb]
    \centering
    \includegraphics[trim={1.8cm 0.75cm 16.0cm 0.8cm}, clip=true, width=0.7\linewidth]{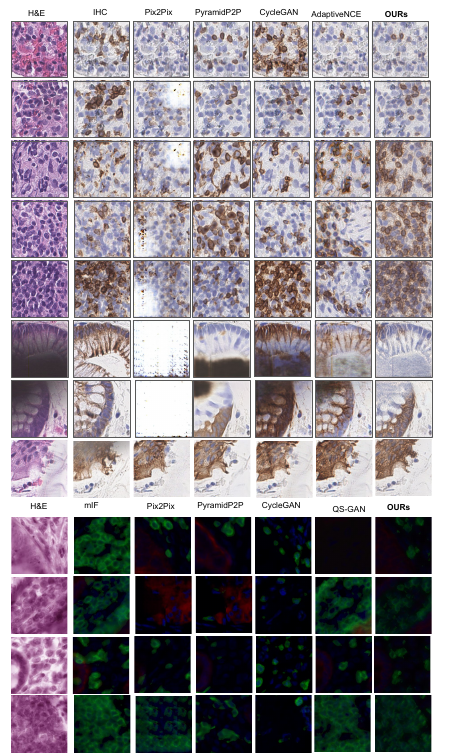}
    \caption{\textbf{Representative failure cases across the CK8/18, CD3, and HEMIT datasets.} Shown are challenging examples for which most methods exhibit reduced performance. In the CD3 dataset, common failure modes include under- and over-staining, potentially reflecting the difficulty of distinguishing CD3-positive cells from morphologically similar lymphoid cells using H\&E appearance alone. For CK8/18, errors are more frequently associated with noisy image regions or reduced tissue quality, occasionally resulting in false-positive staining artifacts. In the HEMIT dataset, failure cases often correspond to samples with atypical color characteristics and weaker visual correspondence between H\&E and mIF channels. These examples highlight challenging scenarios that remain difficult for current virtual staining methods and motivate future work on improving robustness to staining variability, image quality, and dataset-specific inconsistencies.
}
     \label{fig:qualitative_results_add_failure}
\end{figure*}

\begin{figure*}[!htb]
    \centering
    \includegraphics[trim={1.0cm 2.75cm 14.4cm 2.3cm}, clip=true, width=0.9\linewidth]{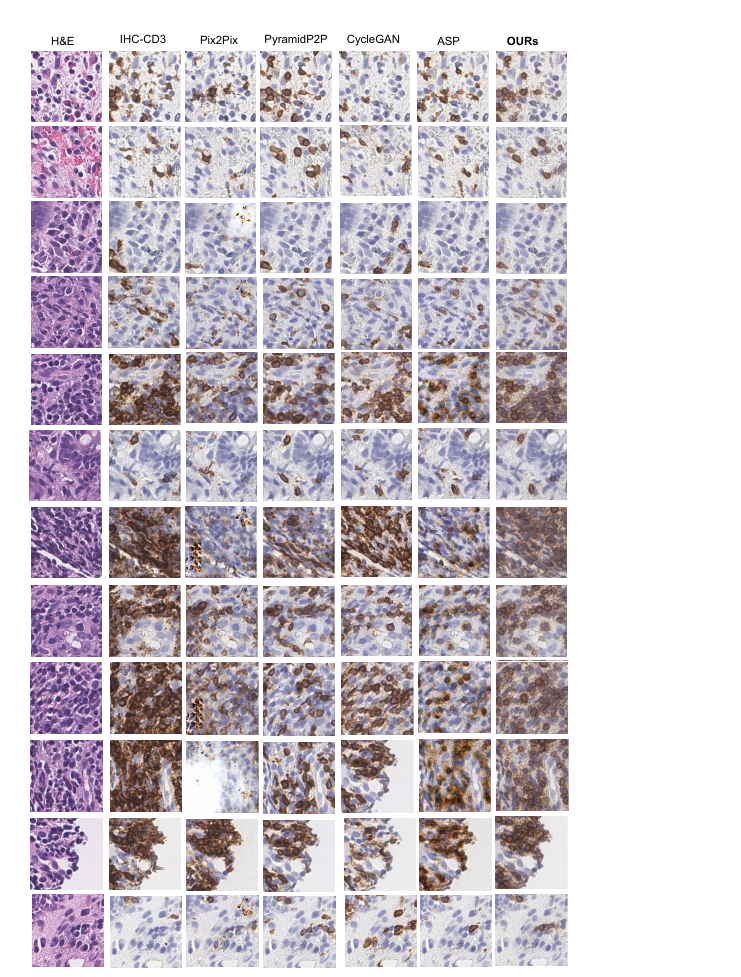}
     \caption{\textbf{ Additional Qualitative Results on CD3 Dataset.}}
     \label{fig:qualitative_results_add_cd3}
\end{figure*}
\begin{figure*}[!htb]
    \centering
    \includegraphics[trim={1.0cm 5.5cm 17.0cm 3.5cm}, clip=true, width=1.0\linewidth]{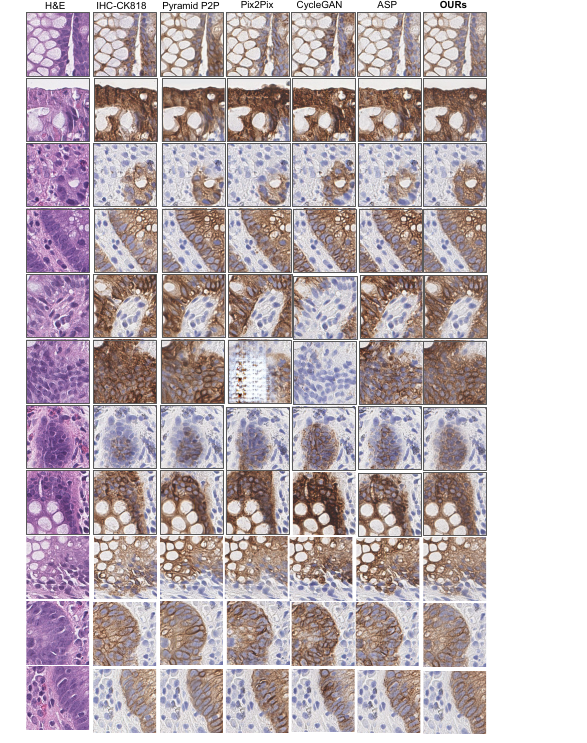}
     \caption{\textbf{ Additional Qualitative Results on CK818 Dataset.}}
     \label{fig:qualitative_results_add_Ck818}
\end{figure*}

\end{document}